\documentclass[12pt,revtex]{article}
\usepackage{graphicx}
\usepackage{amsmath}
\usepackage{amssymb}
\usepackage{amsbsy}
\newcommand{\bit}{\begin{itemize}}
\newcommand{\eit}{\end{itemize}}

\def\benu{\begin{enumerate}}
\def\eenu{\end{enumerate}}

\def\btab{\begin{tabbing}}
\def\etab{\end{tabbing}}

\def\bit{\begin{itemize}}
\def\eit{\end{itemize}}
\def\beq{\begin{equation}}
\def\eeq{\end{equation}}
\def\bec{\begin{center}}
\def\eec{\end{center}}
\def\btable{\begin{tabular}}
\def\etable{\end{tabular}}
\def\beqr{\begin{eqnarray}}
\def\eeqr{\end{eqnarray}}
\def\rarw{\rightarrow}
\def\Rarw{\Rightarrow}

\def\gm{\gamma}
\def\Gm{\Gamma}

\def\eps{\epsilon}

\def\alp{\alpha}
\def\bt{\beta}

\def\dl{\delta}
\def\Dl{\Delta}
\def\sg{\sigma}

\def\rarw{\rightarrow}
\def\del{\partial}

\def\half{\frac{1}{2}}

\def\btab{\begin{tabbing}}
\def\etab{\end{tabbing}}
\def\beqrs{\begin{eqnarray*}}
\def\eeqrs{\end{eqnarray*}}

\def\lan{\langle}
\def\ran{\rangle}

 \fontfamily{}\selectfont
 \usepackage{mathptmx}           

\date{}

\title{Anomalous beam diffusion near beam-beam
synchro-betatron resonances}

\author{Tanaji Sen \\ Accelerator Physics Center\\Fermi National Accelerator
Laboratory \\ Batavia, IL 60510}

\begin{document}

\maketitle


\begin{abstract}
 The diffusion process near low order synchro-betatron resonances driven by
beam-beam interactions at a crossing angle is investigated. Macroscopic
observables such as beam emittance, lifetime and beam profiles are calculated.
These are followed with detailed studies of microscopic quantities such as
the evolution of the variance at several initial transverse amplitudes and
single particle probability distribution functions. We present evidence to
show that the observed diffusion is anomalous and the dynamics follows a
non-Markovian continuous time random walk process. We derive a modified
master equation to replace the Chapman-Kolmogorov equation in action-angle
space and a fractional diffusion equation to describe the density evolution
for this class of processes.
\end{abstract}

\section{Introduction}

Diffusion of particle beams due to nonlinear fields is often a major
source of emittance growth and beam loss in an accelerator. Measurements
of diffusion coefficients have been reported from several hadron 
accelerators \cite{Mess, Fliller, Stancari}. 
The diffusion equation was also used to explain the change in beam
lifetime following the failure of a separator during a Tevatron store
\cite{Sen_2011}. In collision mode the 
beam-beam interactions are usually the dominant nonlinearity.
Diffusion coefficients in the absence of 
low order resonances have been calculated for head-on interactions \cite{Sen96}
and for long-range interactions \cite{Pap_Zim}. Diffusion due to nonlinear
resonances is more complex and the study of this phenomenon has a long
history, see e.g \cite{Hereward, Gerasimov, Shi, Chu, Ohmi07}. Resonances
when modulated, either by dynamical effects such as synchro-betatron
coupling or due to ripple in magnet currents, can sweep across 
phase space and transport particles to large amplitudes \cite{Fisher95}.

In this article we will study the nature of the diffusion process due
to synchro-betatron resonances driven by beam-beam interactions with a 
crossing angle. This was
first investigated at the DORIS collider \cite{Piwinski} and has since
been observed at other colliders. Our aim is to establish the correct
statistical mechanical model that describes the evolution of the beam
density. We examine the possibility that the diffusion process is 
anomalous with detailed tracking simulations and derive a master equation
and a related fractional diffusion equation that may describe the transport
process. A preliminary version of this study was reported in \cite{Sen10}.
An example of anomalous diffusion observed in particle beams as a 
consequence of rf phase modulation was reported in \cite{Jeon}. Anomalous
diffusion processes have been reported in several areas of physics including
plasma turbulence \cite{Negrete05}, and in the motion of laser cooled atoms
on a lattice \cite{Sagi}.

\section{Synchro-betatron resonances due to crossing angles}

Synchro-betatron resonances (SBRs) due to beam-beam interactions at a crossing
angle are 
convenient to study resonantly driven amplitude growth for several 
reasons. At large amplitudes, the
non-linear force vanishes, hence particle excursions do not go to
arbitrarily large amplitudes which is not the case for resonances
due to multipole nonlinearities. This removes numerical instabilities
and also allows the entire beam to be probed for the particle 
dynamics. Another advantage is that the resonances can be studied
in one transverse plane since these resonances are driven by energy
pumped from the longitudinal plane to the transverse plane with 
very little impact on the longitudinal dynamics.

When beams collide at an angle, the transverse distance of a test particle from 
the center of the
opposing bunch depends on the longitudinal position of the particle.
Consequently synchrotron oscillations of the particle couple  to the
transverse beam-beam force leading to excitation of synchro-betatron resonances.
Since the beam-beam force goes to zero at large transverse separations,
the effects of these resonances are experienced by particles only within
a certain range of transverse amplitudes.

For simplicity, we choose the resonances to be in only one transverse
plane, here the horizontal plane. In order to observe effects over
relatively short computation times, we choose low
order resonances. The tunes we choose are unrealistic for operating
colliders but it is likely that the dynamics near high order
resonances is similar but occurs over a longer time scale.

Linear motion and the beam-beam interactions can be described by the
equations of motion resulting from the Hamiltonian
\beq
H = \nu_x J_x + \nu_y J_y + \nu_s J_s + \sum_i^{N_{IP}}U_i(x,y,s)
\dl_P(\phi - \phi_i)
\eeq
where $(\nu_x,\nu_y,\nu_s)$ are the tunes, and $(J_x,J_y,J_s)$ are the
actions. $U(x,y,s)$ is the beam-beam potential, $\dl_P$ is the periodic
delta function, $\phi$ is the azimuthal coordinate and
the sum extends over the number $N_{IP}$ of interaction points. Assuming
Gaussian distributions in all three planes, crossing angles of 
$(2\phi_x,2\phi_y)$ in the horizontal and vertical planes respectively,
the beam-beam potential for colliding proton bunches can be written as
\beqr
U(x,y,s) & = & -\frac{N_b r_p}{\gm_p}\int_0^{\infty} \frac{dq}{[(2\sg_x^2+q) (2\sg_y^2+q)]^{1/2}} \nonumber \\
& & \left( 1 - \exp[-
\frac{(x+s \sin 2\phi_x)^2}{(2\sg_x^2+q)} - \frac{(y+s \sin 2\phi_y)^2}{(2\sg_y^2+q)}] \right)
\eeqr
where $N_b$ is the bunch intensity of the opposing bunch, $r_p$ is
the classical proton radius, $\gm_p$ is the proton energy in units of
its rest mass and $\sg_x, \sg_y$ are the rms beams sizes of the opposing
beam at the interaction point (IP). The potential can be expanded as
a Fourier series
\beq
U(x,y,s) = \sum_{m_x,m_y,m_s,p} U_{m_x, m_y, m_s} \exp[i(m_x\psi_x+m_y\psi_y
 + m_s\psi_s - p\phi)]
\eeq
This potential can excite 
synchro-betatron resonances given by the resonance condition
$m_x \nu_x + m_y\nu_y + m_s\nu_s = p$ where $(m_x,m_y,m_s,p)$ are
integers. It can be shown from the structure of the Fourier harmonics
$U_{m_x, m_y, m_s}$ that they are non-zero only when the sum
$m_x+m_y+m_s$ is even. The Fourier harmonics can also be
used to calculate the tune shifts with amplitude and the resonance
driving terms, as was done in \cite{Sen_TeV}. As one example, we write
down the zero transverse amplitude tune shift for round beams. This 
tune shift now depends on the longitudinal oscillation amplitude $a_s\sg_s$
as
\beq
\Dl\nu_x(a_x=0,a_y=0,a_s) = \xi e^{-\tau}[I_0(\tau) - I_1(\tau)(1+ \half \frac{(a_s h_x)^2}{\tau})]
\eeq
and a similar expression for $\Dl\nu_y$. Here
$\xi=N_b r_p/(4 \pi \eps_N)$ is the usual beam-beam parameter, 
$(a_x\sg_x, a_y\sg_y)$ are the transverse amplitudes of the particle,
$I_0, I_1$ are 
modified Bessel functions and the other dimensionless parameters are
\[ 
\tau = \frac{1}{4}a_s^2(h_x^2+h_y^2), \;\;\; 
h_x = \frac{\sg_s}{\sg_x}\sin 2\phi_x, \;\;\; 
h_y = \frac{\sg_s}{\sg_y}\sin 2\phi_y
\]
As a consequence, only those zero transverse amplitude particles with zero
longitudinal amplitude $a_s$ experience the full beam-beam tune shift $\xi$.
Particles with non-zero amplitude $a_s$ experience a smaller tune shift.

Since the LHC employs crossing angles in its collision scheme, we will
use the LHC beam parameters in the simulations reported here. As in
the LHC, the crossing angle is in the horizontal plane at one IP and
in the vertical plane at the second IP. We consider resonances excited
in the horizontal plane only, so they are of the form 
$m_x \nu_x + m_s\nu_s = p$ with $m_x + m_s$ even. In our model
the only sources of tune spread are the beam-beam interactions. These 
interactions between protons lowers the betatron tunes at small amplitudes.
We choose the large amplitude tunes, i.e. the tunes with only the linear 
lattice, to satisfy one of the SBR resonance conditions. 
Having chosen a particular resonance $m_x\nu_x + mu_s\nu_s = p$ to be
satisfied by the bare lattice tunes, the tunes inside the bunch are 
determined by the beam-beam parameter $\xi$, the synchrotron tune $\nu_s$ and
the amplitudes $(a_x,a_y, a_s)$ of the particle.
The nominal LHC
horizontal tune is 0.31 at collision, so we searched among the following 
resonances: $3\nu_x \pm \nu_s = 1$, $2(3\nu_x \pm 2\nu_s) = 2$ as well as
$2(4\nu_x \pm \nu_s) = 2$ and $4\nu_x \pm 2\nu_s = 1$ to find those
that cause large growth of the emittance and beam tails. 
Given that the betatron tune spread from head-on beam-beam interactions is 
about 0.007 and the small 
amplitude synchrotron tune is $\sim$0.002, the choices $2(3\nu_x - 2\nu_s) = 2$ and $4\nu_x - 2\nu_s = 1$ had the greatest impact on the beam.
With these choices, low amplitude particles are resonant with the 
third and fourth order betatron resonances respectively, and the synchrotron
oscillations modulate these resonances leading to large amplitude
growth. The other resonances are resonant at larger amplitudes and 
consequently have a smaller impact on the bunch. 
The bare lattice (which become the large amplitude) betatron tunes corresponding
to these resonances are shown in Table \ref{table: param}. Some of these
parameters may be slightly different from the present LHC design values,
e.g the LHC design value of the crossing angle is 285$\mu$rad.

\begin{table}
\caption{Table of basic parameters in simulation model.
Resonance I is $2(3\nu_x - 2\nu_s) = 2$, resonance II is 
$4\nu_x-2\nu_s=1$.}
\bec
\btable{|c|c|} \hline
Beam parameter & Value \\ \hline
Energy [TeV] & 7.0 \\
Bunch Intensity & 1.1$\times 10^{11}$ \\
$\sg_x, \sg_y$ [$\mu$m] & 16.6, 16.6 \\
$\sg_s$ [cm] & 7.5 \\
Rf voltage [MV] & 16 \\
Crossing angles [$\mu$rad] & 300 \\
Beam-beam parameter & 0.0034 \\
Resonance I: $(\nu_x,\nu_y)$ & 0.3353, 0.32 \\
Resonance II: $(\nu_x,\nu_y)$ & 0.2514, 0.32 \\
\hline
\etable
\eec
\label{table: param}
\end{table}

\section{Simulations of beam variables}

In this section we will describe multi-particle simulation results.
These will include the  emittance growth, evolution of beam profiles,
amplitude growth at different initial amplitudes,
and also the growth of the variance in action at these initial amplitudes.
This will allow us to probe both the macroscopic and microscopic beam
behaviour.

The simulations were performed with a simple numerical model 
consisting of six dimensional linear transport 
between the two collision points, a sinusoidal 
longitudinal map through an rf cavity and weak-strong beam-beam 
interactions at the two IPs. 
The beam-beam interactions occur with a horizontal crossing angle at one
IP and a vertical crossing angle at the second IP. 
The strong beam was assumed to have a Gaussian distribution in all three planes.
Magnetic nonlinearities 
are not included, both to keep the model as simple as possible and also to
avoid particle amplitudes from growing exponentially fast far from the beam core.
Limiting amplitude growth to finite values allows us to keep all particles in the
distribution and hence study the growth of the beam tails with good statistics.

\subsection{Emittance growth and lifetimes}

Emittance growth was calculated by evolving ensembles of $N$ particles
(5000 $ \le N \le $ 20000) starting with Gaussian distributions in
all planes. Typically 10,000 particles sufficed to obtain results that
did not change much with a larger number of particles. 
The calculated emittance was the rms emittance, e.g.
$\eps_x = [\lan x^2 \ran \lan x^{'2}\ran - \lan x x'\ran^2]^{1/2}$.
\begin{figure}
\centering
\includegraphics[scale=0.55]{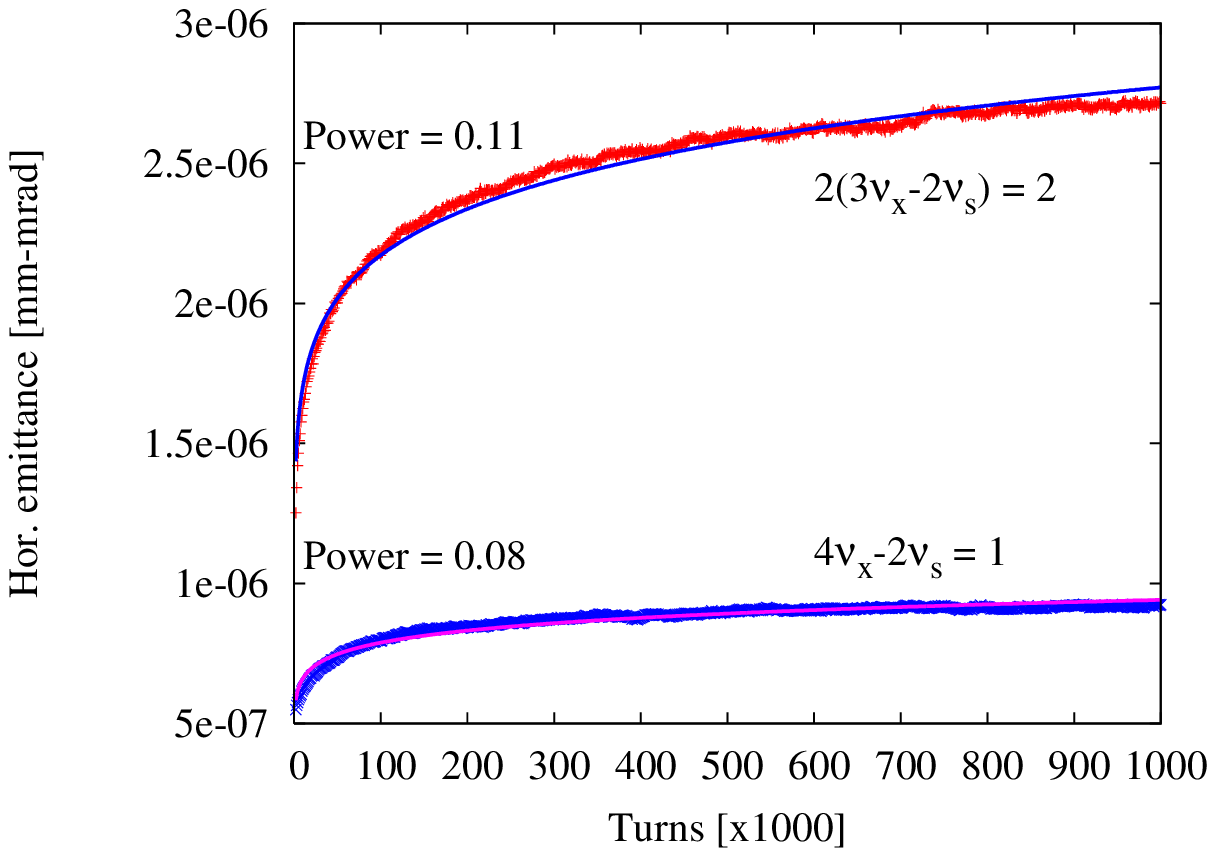}
\includegraphics[scale=0.55]{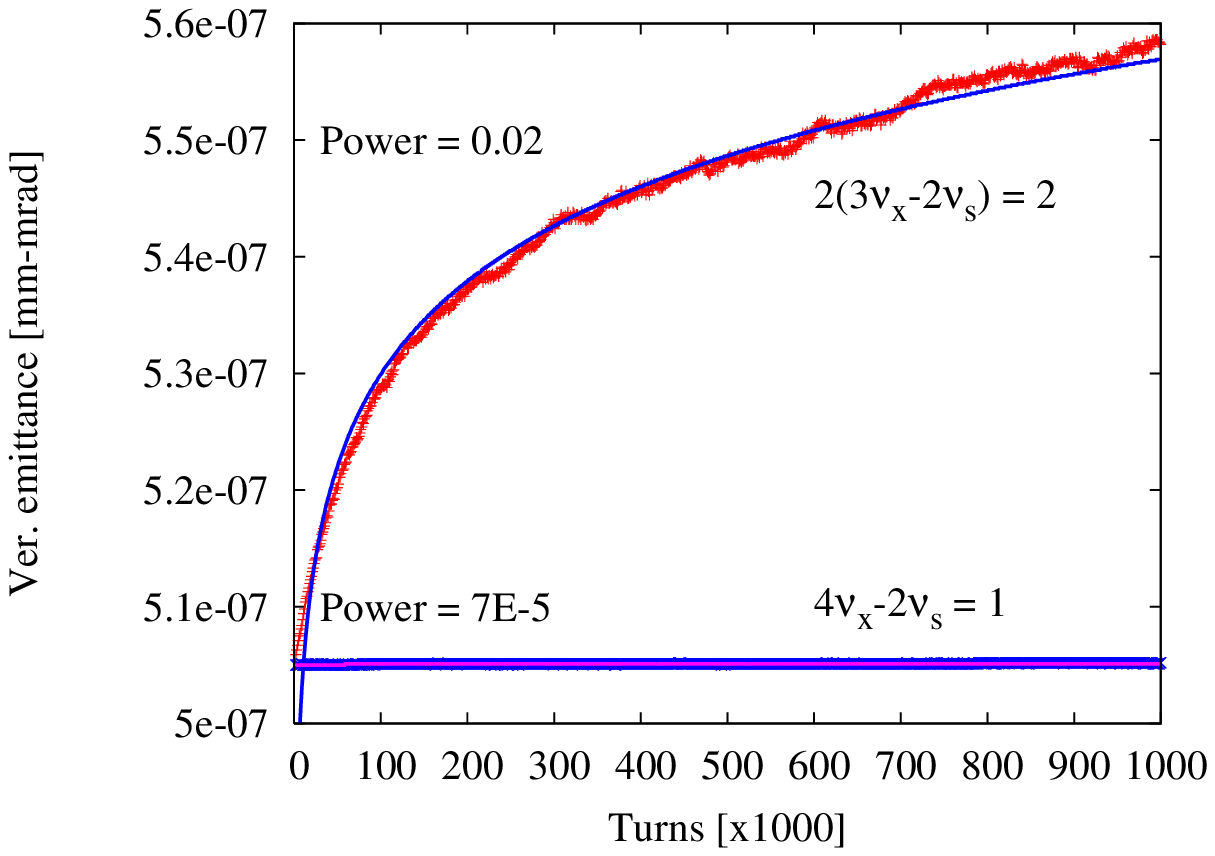}
\caption{(color) Emittance growth vs turns with tunes on the synchro-betatron resonances.
Left: horizontal emittance growth, Right: vertical emittance growth. The power
law fits and the exponents for the fits are also shown.}
\label{fig: emittances}
\end{figure}
Figure \ref{fig: emittances} shows the emittance growth with 20,000 particles
on the two resonances. We find that the growth follows a simple power law,
the fits are also shown in the figure. 
We observe that the horizontal emittance growth after 10$^6$ turns is more than
2.5 times larger on the $2(3\nu_x-2\nu_s)=2$ resonance than on the
 $4\nu_x-2\nu_s=1$ resonance. The vertical emittance growth is much smaller
than the horizontal, about a factor of five smaller for the first resonance
and it is practically zero for the second resonance.

\begin{figure}
\centering
\includegraphics[scale=0.5]{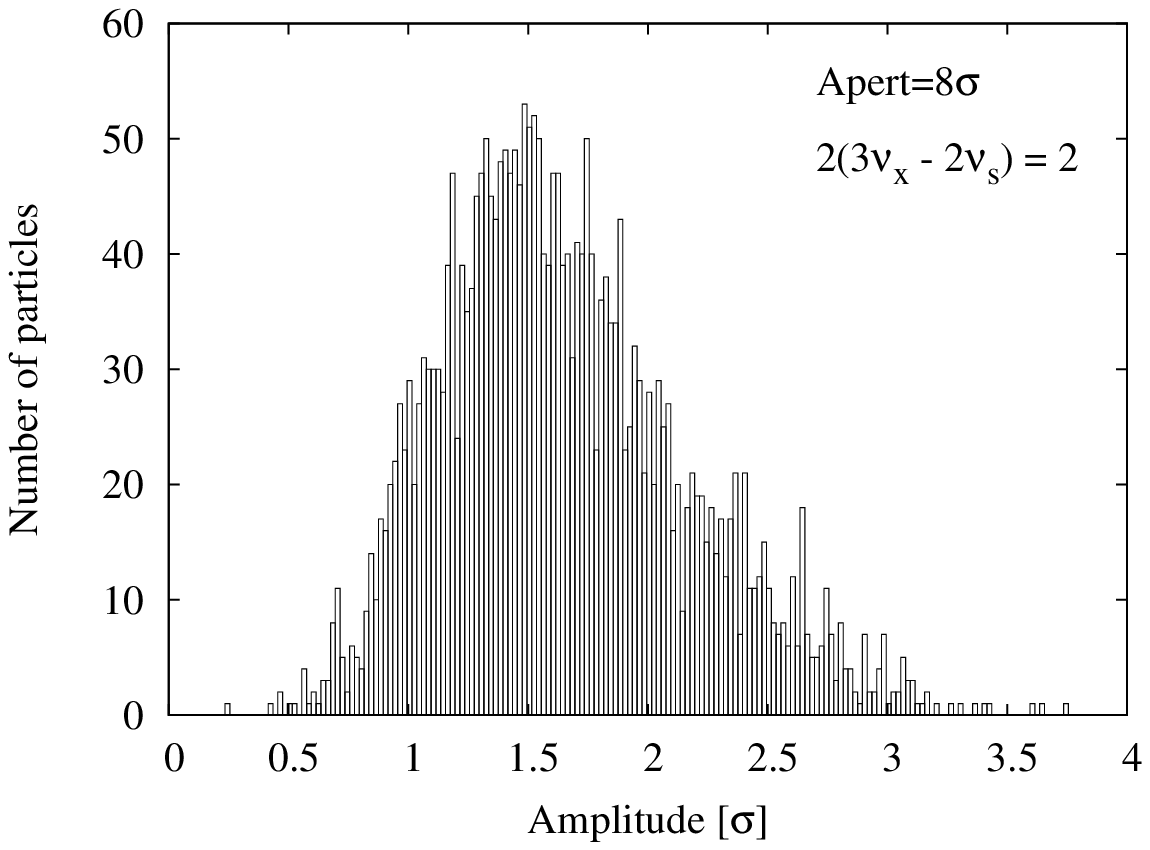}
\includegraphics[scale=0.5]{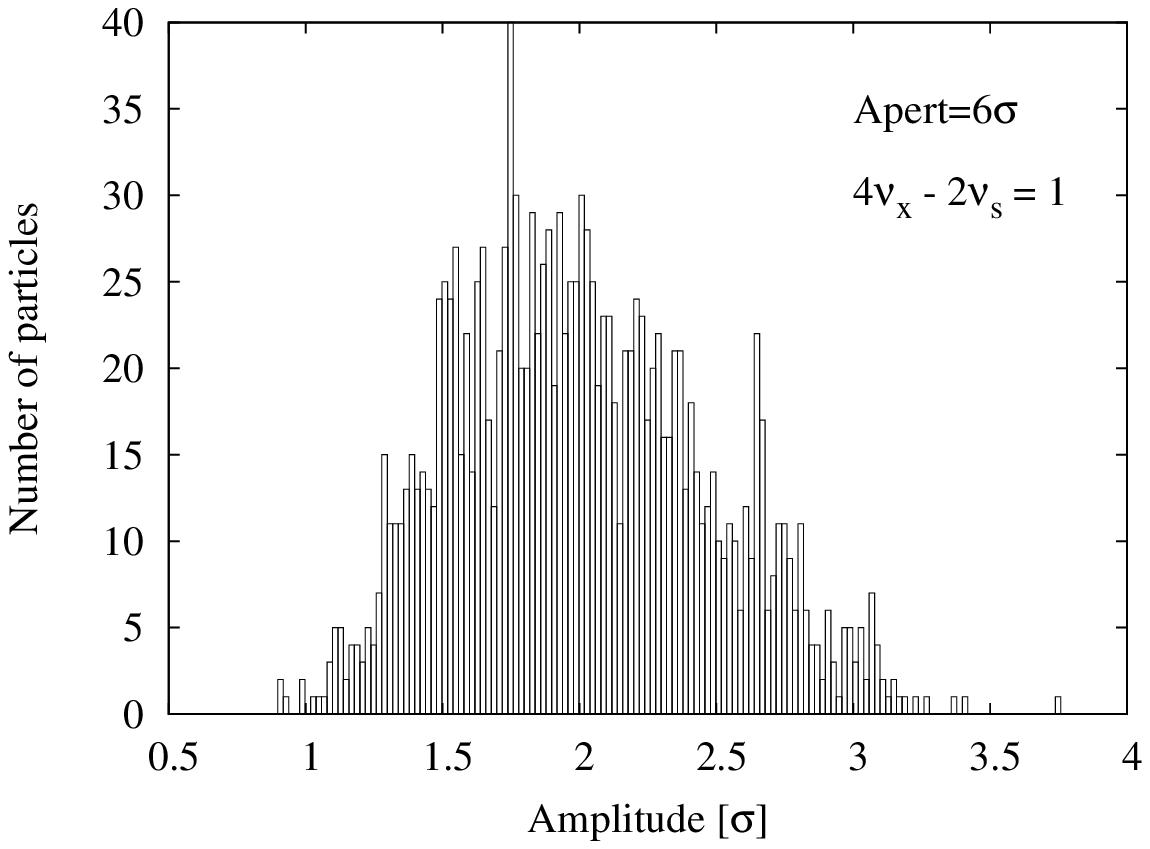}
\includegraphics[scale=0.5]{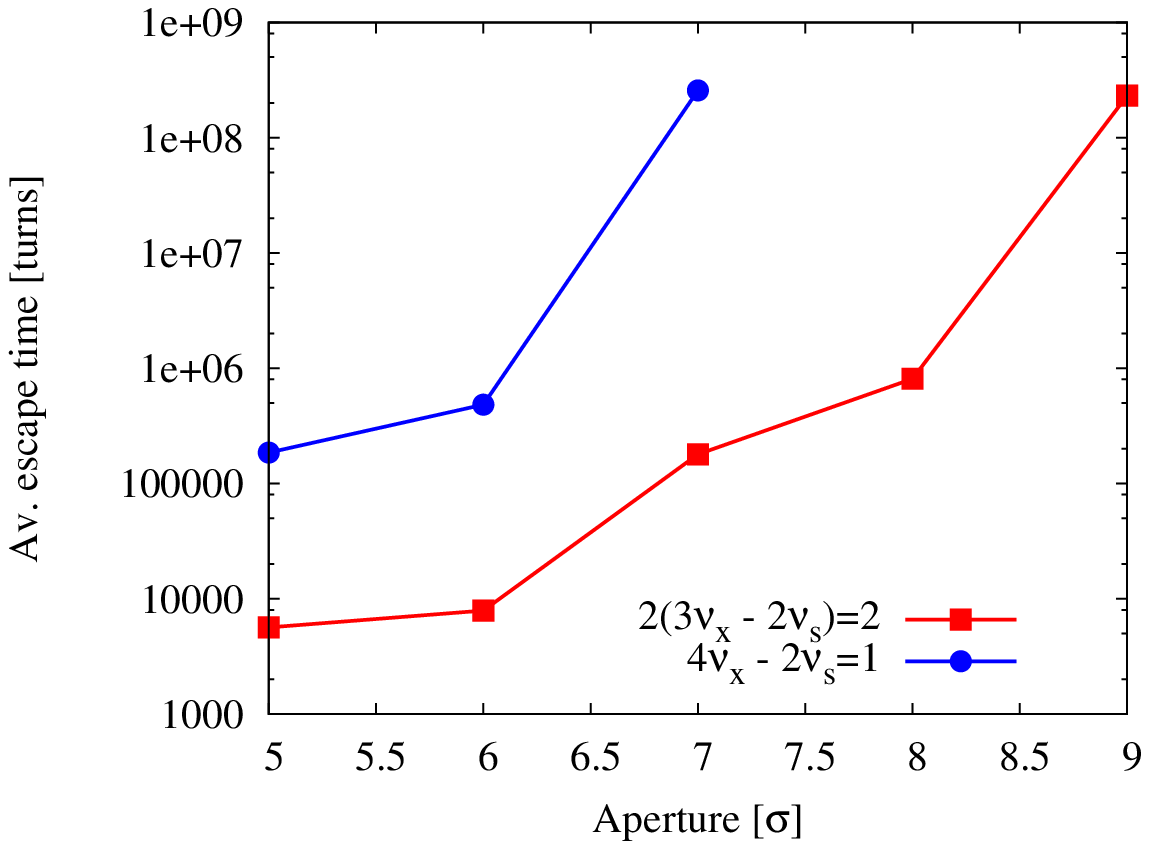}
\caption{Top: Distribution of amplitudes reaching an aperture of 8$\sg$ on the
$2(3\nu_x-2\nu_s)=2$ resonance (left) and an aperture of 6$\sg$ on the
$4\nu_x-2\nu_s=1$ resonance (right). In both cases, the initial distribution
was a Gaussian with 40,000 particles.
Bottom (color): Average escape time for the two resonances at different apertures.}
\label{fig: escapetimes}
\end{figure}
By imposing a finite aperture restriction, we can find the escape time needed by
particles to reach this aperture. This has been calculated for several different
apertures and for both resonances. Apertures were placed from 5$\sg$ to 10$\sg$
at intervals of 1$\sg$. On the $2(3\nu_x-2\nu_s)=2$ resonance, we find
that about 7\% of particles reach 8$\sg$, a handful reach 9$\sg$ and none
reach 10$\sg$. On the $4\nu_x-2\nu_s=1$ resonance, about 4\% of particles
reach 6$\sg$, a few reach 7$\sg$ and none reach 8$\sg$. The amplitude 
distribution of the particles reaching 8$\sg$ on the first 
resonance and of the particles reaching 6$\sg$ on the second
resonance are shown in the top plots of Fig \ref{fig: escapetimes}. The initial 
distribution
in each case was a Gaussian with 40,000 particles. On the $2(3\nu_x-2\nu_s)=2$ 
resonance, the maximum of the amplitude distribution occurs close to 1.5$\sg$ -
an amplitude close to the lower edge of the resonance islands, shown later in
Fig \ref{fig: xpspace_3rd}. The minimum amplitude that reaches the aperture is
0.25$\sg$. 
On the $4\nu_x-2\nu_s=1$ resonance, the 
corresponding peak in the amplitude distribution is close to 1.8$\sg$, also at
the lower edge of the resonance islands seen in Fig. \ref{fig: xpspace_4th}.
The minimum amplitude that reaches the aperture on this resonance is 0.9$\sg$.

 The average escape time in the simulation may be interpreted as representing
the beam lifetime. 
The bottom plot in Fig. \ref{fig: escapetimes} shows the
average escape time (calculated with 40,000 particles) as a function of the 
aperture amplitude for both resonances. The average escape time with 20,000
particles yielded similar values showing that these numbers have converged to
stable values. The average escape time increases by an order of magnitude or more
for each increase in aperture by 1$\sg$. The average escape time at 8$\sg$ on the
$2(3\nu_x-2\nu_s)=2$ resonance is about the same as at 6$\sg$ on the
$4\nu_x-2\nu_s=1$ resonance. At a fixed aperture, 
the differences in escape times between the
two resonances increases by about two orders of magnitude at 5 and 6 $\sg$ and 
three orders of magnitude at 7$\sg$. One would expect this trend of increasing
lifetimes to continue with higher order resonances.

\subsection{Beam profiles}

\begin{figure}
\centering
\includegraphics[scale=0.55]{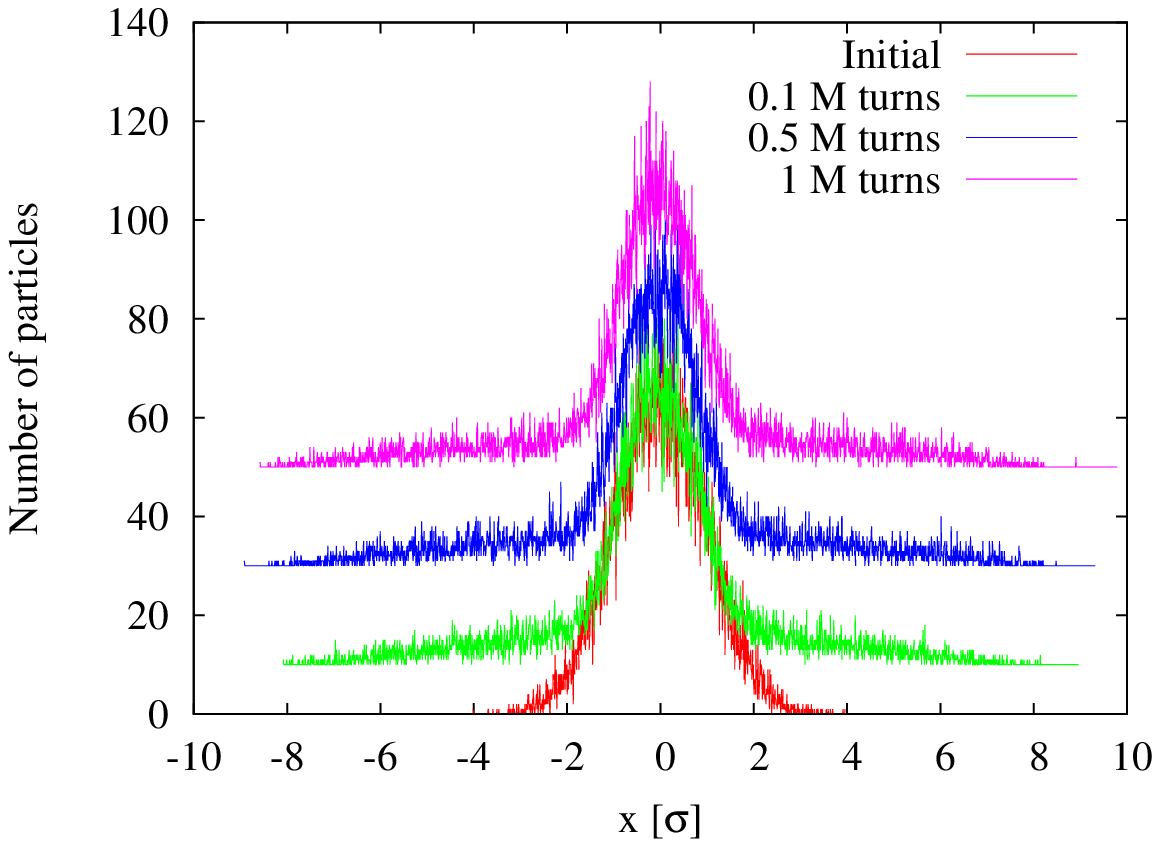}
\includegraphics[scale=0.55]{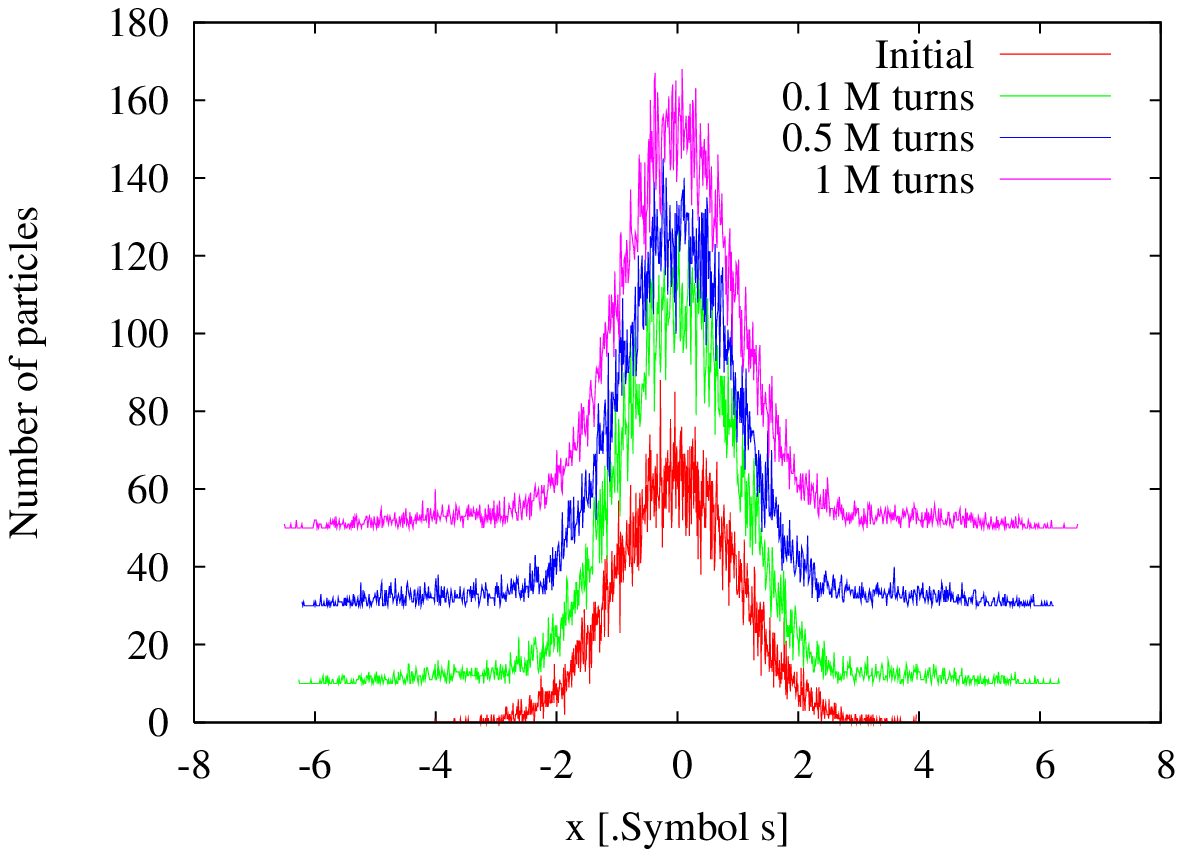}
\caption{(color) Mountain range view of the horizontal beam profile 
initially and at subsequent times. Particles were on the resonance 
$2(3 \nu_x - 2 \nu_s) = 2$ (left) and on the resonance $4 \nu_x - 2 \nu_s = 1$
(right).}
\label{fig: pdfx_3rd_4th}
\end{figure}

 The beam profiles were found for the same distributions and resonances. 
The left plot in Fig \ref{fig: pdfx_3rd_4th} shows a mountain range view 
of the horizontal beam profiles (i.e. distribution function of the
horizontal position), initially and then at other intervals
up to 10$^6$ turns with tunes on the resonance $2(3\nu_x - 2\nu_s)=2$.
After the initial time, the subsequent horizontal profiles develop
long non-Gaussian tails which extend out to $\tilde \pm 8\sg$ compared
to the initial Gaussian distribution which was limited to $\pm 3.5\sg$. The
vertical beam profiles (not shown here) however stayed Gaussian and
close to the initial distribution. The right plot in this figure
shows the horizontal profiles but with tunes on resonance $4\nu_x-2\nu_s=1$.
We observe that in this case as well that the tails are non-Gaussian
and extend out to about $\pm 6\sg$, not quite as far as on the 
first resonance. Again there is very little change in the vertical profile.
We observe that the beam tails do not appear to change very much after the
first 100,000 turns or so.
It is most likely that the regions of enhanced diffusion are depleted
within these turns. The particles in the vicinity of
the resonance islands are transported to larger amplitudes quickly and are 
detuned from the resonance. The amplitudes to which they move have
much smaller diffusion, so the beam tails do not change much. As we will see
in the next subsection, the evolution in the beam core 
shows growth even after several hundred thousand
turns. However these particles do not migrate to the tails
during the time duration followed. Thus we continue to observe emittance growth.

In order to find distributions that can best fit the non-Gaussian tails,
we first look to the Central Limit Theorem (CLT) which explains the ubiquity
of the Gaussian distribution. 
This powerful theorem states that the distribution of a sum 
of a sequence of random, identically distributed and independent 
variable  with finite mean and second moment tends to a Gaussian
distribution in the limit that the number in the sequence approaches
infinity. Generalizing the CLT by dropping the requirement of a finite
second moment leads to the family of Levy stable distributions 
\cite{Levy-stable}. For applications in beam dynamics, these distributions 
will still have a finite second moment because they do not extend to 
infinity but are truncated at the beam pipe or the closest physical apertures. 

Levy stable distribution functions are defined by an inverse Fourier 
transform of a stretched exponentially decaying function in Fourier
space
\beq
L_{\alp}(z) = \frac{1}{2\pi}\int_{-\infty}^{\infty}
\exp[-i z k - |k|^{\alp}] dk, \;\;\;\; 0 < \alp < 2
\eeq
There is no known closed form expression for arbitrary values of $\alp$.
Special cases include: the Lorentz distribution $L_{1}(z)$ while
$L_2(z)$ is the Gaussian distribution. There are more general 
asymmetric versions of the Levy stable distribution with additional parameters 
but we shall not need them here. Some basic
properties of these functions are \cite{Montroll}
\bit
\item These functions are normalized : 
$ \int_{-\infty}^{\infty} dz L_{\alp}(z) =  1$

\item They are even functions : 
$ L_{\alp}(-z) = L_{\alp}(z) $

\item At $z=0$, 
$ L_{\alp}(0) = \frac{1}{\pi \alp} \Gamma(\frac{1}{\alp}) $, 
which increases rapidly when $\alp \rarw 0$. 

\item At large values of $z$, the distributions decay as
\[
\lim_{z\rarw\infty} L_{\alp}(z) \sim \frac{1}{\pi}\sin(\half\pi \alp)
\frac{\Gamma(1+\alp)}{|z|^{1+\alp}}
\]

\eit

\begin{figure}
\centering
\includegraphics[scale=0.55]{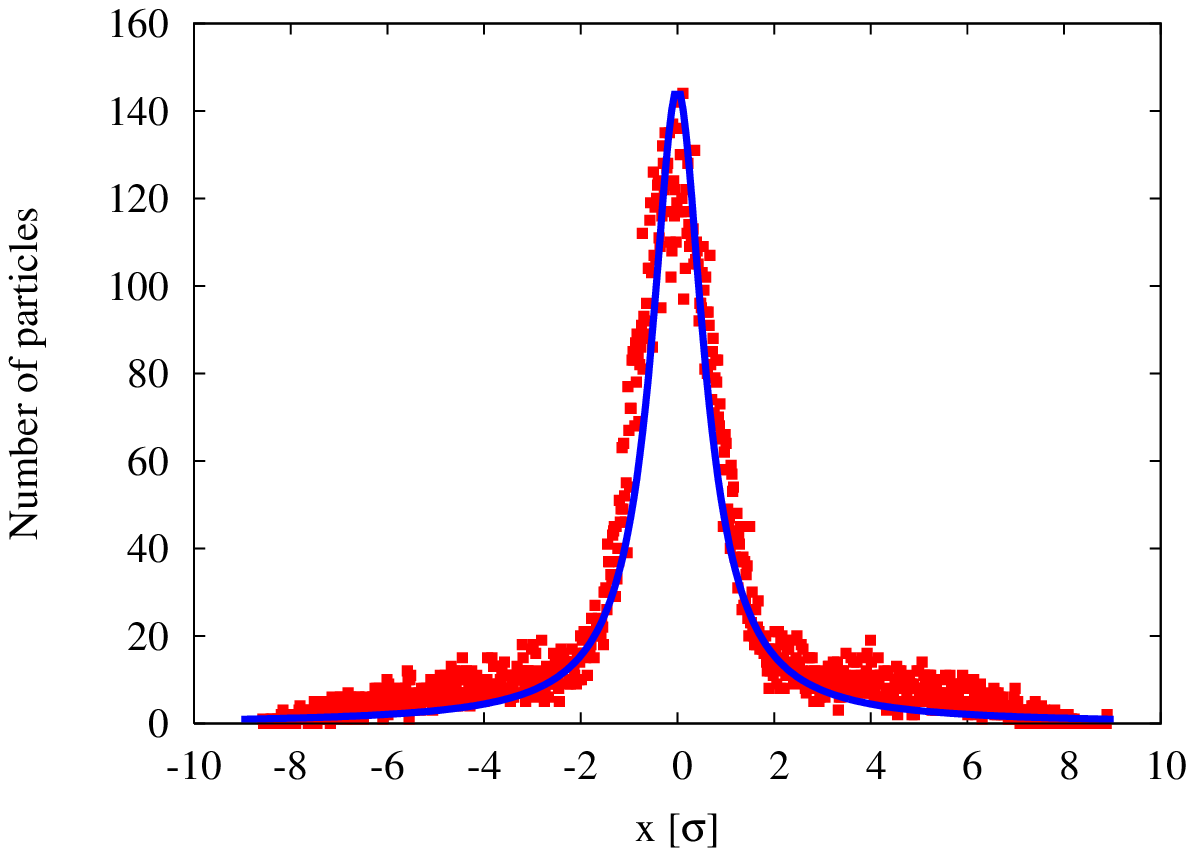}
\includegraphics[scale=0.55]{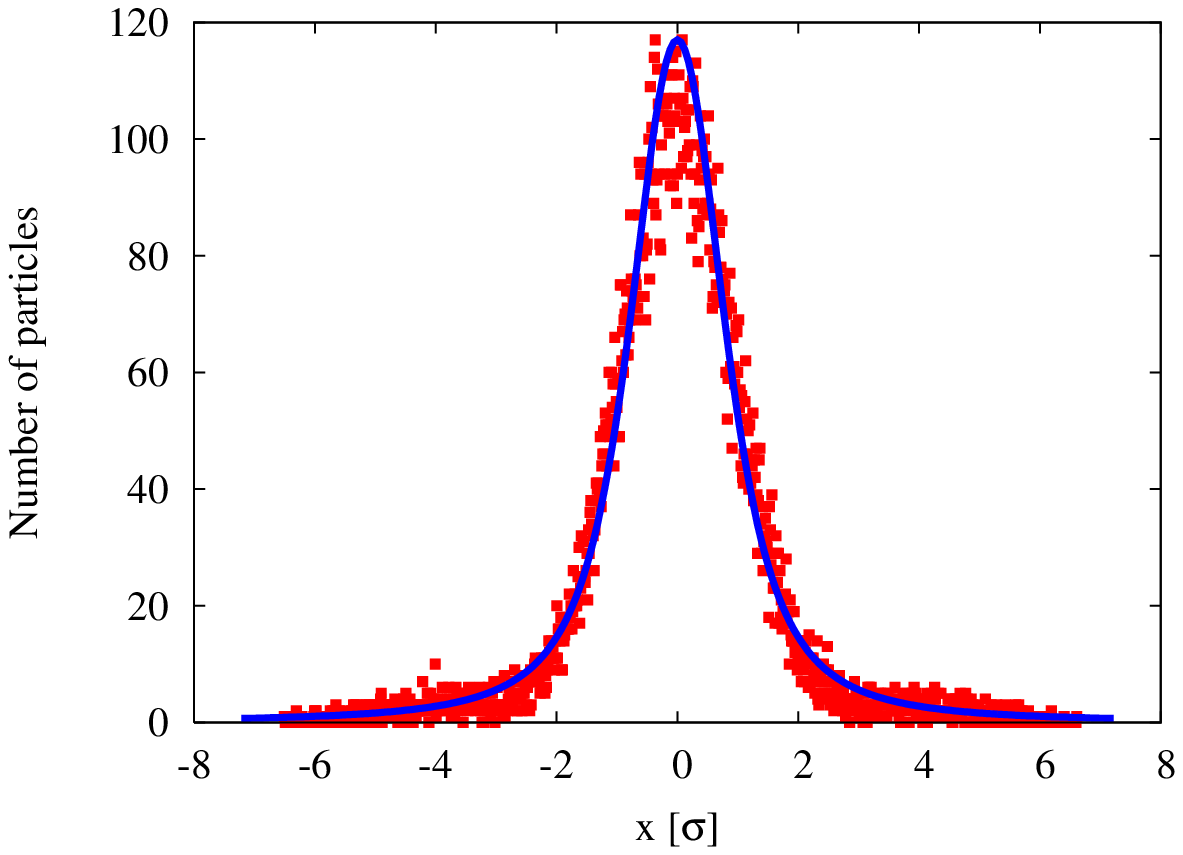}
\caption{(color) The final horizontal profile and a fit (blue) with a Levy stable 
distribution $L_{\alp}$. Left: Resonance $2(3 \nu_x - 2 \nu_s) = 2$ and 
$\alp=0.95$. Right: Resonance $4 \nu_x - 2 \nu_s = 1$ and $\alp=1.3$.}
\label{fig: pdfx_Levyfit}
\end{figure}

We find that the non-Gaussian 
horizontal profiles can be fit by these Levy stable distributions $L_{\alp}$. 
The left plot in Fig. \ref{fig: pdfx_Levyfit}
shows the fit of the final horizontal profile for the resonance
$2(3 \nu_x - 2 \nu_s) = 2$ with a Levy stable
distribution with parameter $\alp=0.95$. This profile is narrower than a 
Lorentzian and decays at large $x$ as $|x|^{-1.95}$.
The right plot in this figure shows the final distribution on the resonance
$4 \nu_x - 2 \nu_s = 1$ can also be 
fit by a Levy stable distribution with a larger central width and
corresponding to $\alp=1.3$. This profile is wider than a 
Lorentzian and decays at large $x$ as $|x|^{-2.3}$.
The Levy stable distributions were generated with a Mathematica 
package \cite{Rimmer_Nolan}.

It is known \cite{Negrete03} that the Levy stable distributions serve
as Green's functions to fractional diffusion equations for a density
$\rho(x,t)$ of the type
\beq
\frac{\del}{\del t}\rho(x,t) = \chi \; \mbox{}_{-\infty}D_x^{\alp}\rho(x,t)
\label{eq: Levy_diffeq}
\eeq
where $\chi$ is a constant diffusion coefficient and 
$\mbox{}_{-\infty}D_x^{\alp}$ is the Riemann-Liouville fractional
space derivative of order $\alp$ given by,
\beq
 \mbox{}_{-\infty}D_x^{\alp}\rho = \frac{1}{\Gm(2-\alp)}\frac{\del^2}{\del x^2}
 \int_{-\infty}^x \frac{\rho(x')}{(x - x')^{\alp-1}} dx'
\eeq
The solution of the fractional diffusion equation above is
\beq
\rho(x,t) = \int_{-\infty}^{\infty} L_{\alp}(z) \rho_0(x- (\chi t)^{1/\alp}z) dz
\eeq
where $\rho_0(x)$ is the initial density. Levy stable distributions have also
been shown to be solutions of other fractional diffusion equations \cite{West}.
There is no reason to believe that either
Equation (\ref{eq: Levy_diffeq}) or of the type in reference \cite{West}
are appropriate for our problem. However the fact that the long time beam profiles are described by these
Levy distributions is our first indication that the amplitude growth 
process may be described by an appropriate fractional diffusion equation
rather than the regular diffusion equation.
In Appendix A we derive a different fractional diffusion equation that may
describe the dynamics observed here.

\subsection{Growth at individual amplitudes}

We now take a closer look inside the beam distribution to determine
how the amplitude growth changes with amplitude.
Instead of a Gaussian distribution in phase space, we consider
delta function distributions in action.
We select a discrete number of horizontal actions and at
each action we place 4000 particles uniformly distributed in angle.
The vertical amplitude was kept constant at 0.1$\sg$ for all particles.
The initial distribution in transverse action angle space can be written as
\beq
\rho(J_x,\theta_x,J_y,\theta_y) = \dl(J_y - J_{0.1})
P(\theta_x) P(\theta_y) \sum_i \dl(J_x - J_i)
\eeq
where $J_{0.1}$ is the action at an amplitude of 0.1$\sg$, 
$P(\theta_x)$ is a uniform distribution in the horizontal angles etc.
The initial longitudinal variables were chosen to be the same for all
particles: $z = 1\sg_s$, $\dl p/p = 1 \sg_p$ in these simulations.
We let these distributions evolve and record the final distribution in
amplitude after 10$^6$ turns. 
\begin{figure}
\centering
\includegraphics[scale=0.25,angle=-90]{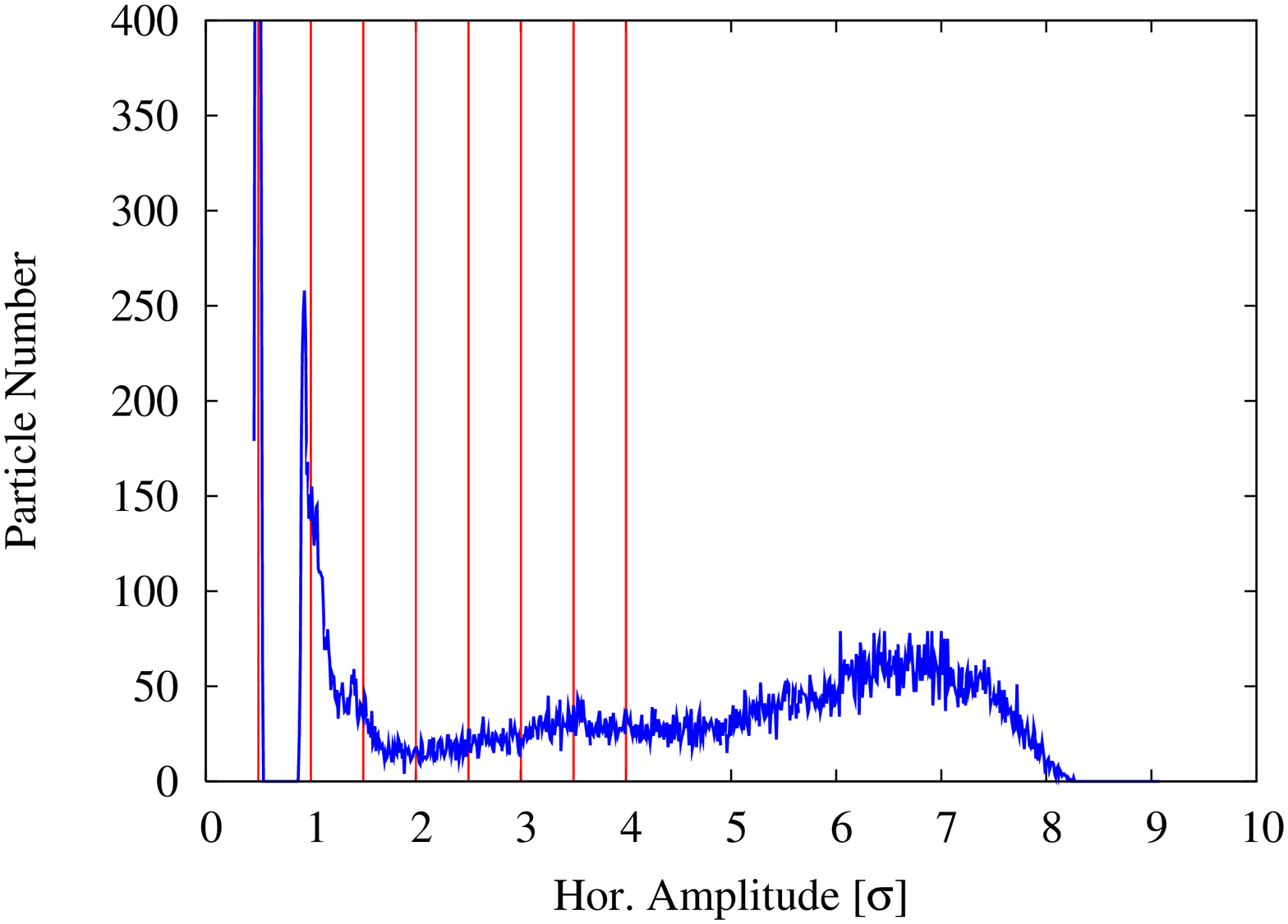}
\includegraphics[scale=0.25,angle=-90]{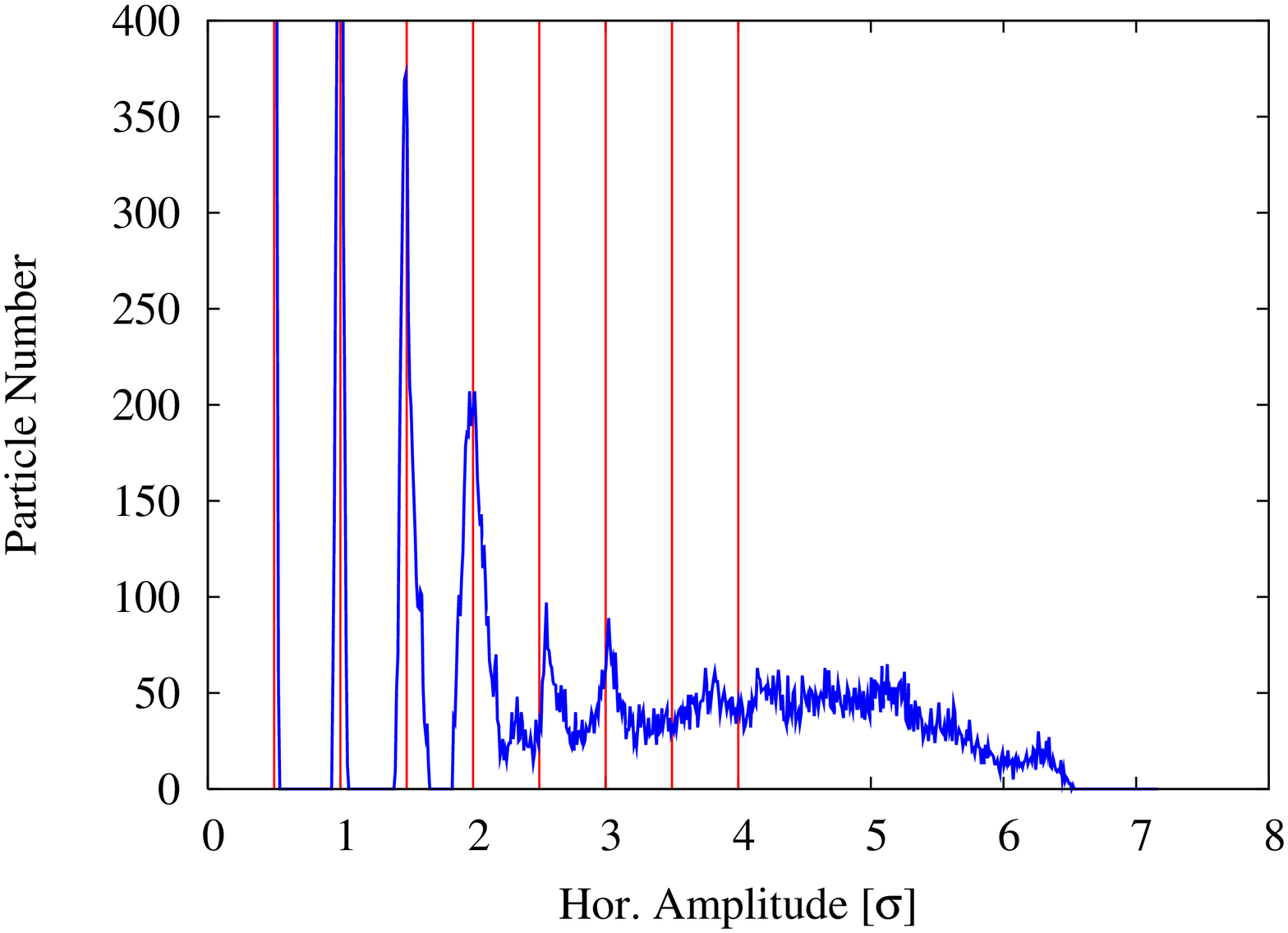}
\caption{(color) Plots of the initial (red) and final (blue) beam distributions 
in the horizontal plane at the resonance $2(3 \nu_x - 2 \nu_s) = 2$ (left) and 
resonance $4 \nu_x - 2 \nu_s = 1$ (right) respectively.
Initially 4000 particles each were placed  at horizontal amplitudes
from 0.5 to 4$\sg$ in steps of 0.5$\sg$. 
The vertical amplitude was kept constant at 0.1$\sg$. In the plots the
vertical scale has been truncated to 400 in order to show clearly the particle 
numbers in the final distribution at large amplitudes.}
\label{fig: ampx_distribs}
\end{figure}
The left plot in Fig \ref{fig: ampx_distribs} shows the initial (red)
and final (blue) distributions for resonance I. We observe that 
particles at 0.5$\sg$ stay close to their initial amplitude. At 1$\sg$,
many particles have moved to larger amplitudes but a sizable fraction
stay in their original neighbourhood. This shows a large variation in
final amplitude depending on their initial angle or sensitivity to
their initial conditions. It suggests that motion in the neighbourhood
of 1$\sg$ could correspond to bounded chaos. At amplitudes of 1.5$\sg$
and higher, the vast majority of particles have migrated to larger
amplitudes up to 8$\sg$ and depleted the initially populated regions.
There is a broad local maxima in the final
distributions at $\sim 7\sg$.  The right plot in Fig. 
\ref{fig: ampx_distribs} shows the corresponding results for resonance
II. The results are qualitatively similar with some differences. 
The initial amplitude with large variation in final amplitude is closer
to 2$\sg$ and the largest amplitude reached is about 7$\sg$. 
On this resonance there remain local spikes at 2.5 and 3$\sg$ showing that
diffusion at these amplitudes is weaker than in the first resonance.

\subsection{Variance of the action and diffusion type}

\begin{figure}
\centering
\includegraphics[scale=0.25,angle=-90]{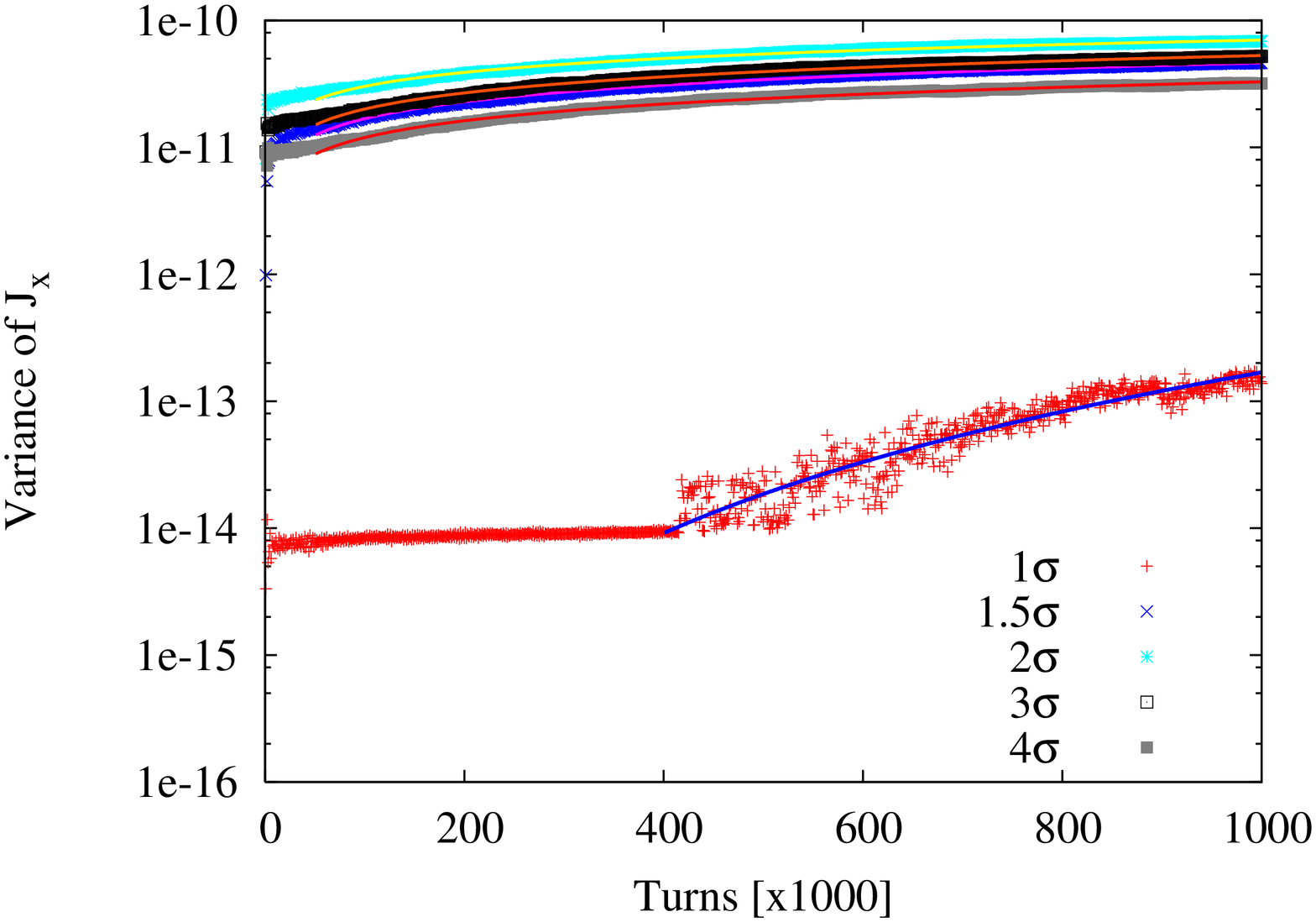}
\includegraphics[scale=0.25,angle=-90]{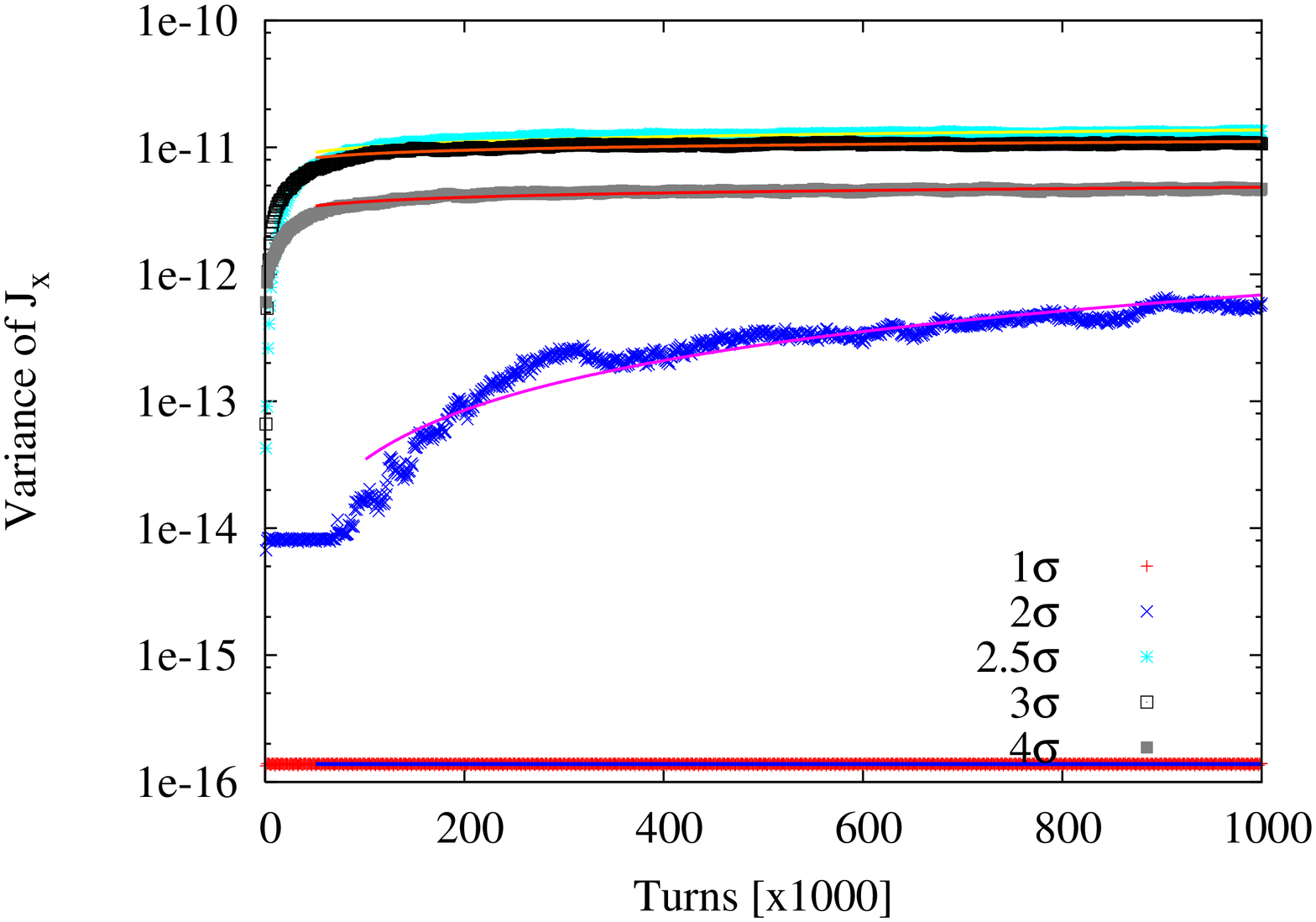}
\caption{(color) Variance in the horizontal actions over time at tunes 
corresponding to the resonances $2(3 \nu_x - 2 \nu_s) = 2$(left) and
$4\nu_x - 2 \nu_s = 1$ (right). Also shown are monomial fits to the data.
Note that the variances are plotted on a logarithmic scale. All variances
are zero initially but the zero is suppressed here. }
\label{fig: varianceJx}
\end{figure}

We now examine the diffusion from individual amplitudes. In
regular diffusion the variance of the diffusing quantity, here the action,
grows linearly with time which allows one to define time independent 
diffusion coefficients $D(J) = \lan (\Dl J)^2 \ran/\Dl t$. We check the 
validity of this assumption for the beam-beam driven SBRs. 
using the same initial distributions as used in Fig \ref{fig: ampx_distribs}.
Variances are calculated over particles at the same initial 
action. Figure
\ref{fig: varianceJx} shows the growth in the variance of the horizontal
action at several initial actions for both resonances. The vertical 
amplitude was constant at $y=0.1 \sg$.
Initially the variance is zero at all actions
but then grows at different rates depending on the action. The growth in the
variance is not linear at any action. In most cases there is a sharp initial 
transient growth which is followed by a slower long term growth. This long
term growth can be modeled (again in most cases) by a power law behavior of
the form
\beq
\lan (\Dl J_x)^2 \ran \sim C_x t^{p_x}, \;\;\;
\lan (\Dl J_y)^2 \ran \sim C_y t^{p_y}
\eeq
where the coefficients $(C_x,C_y)$ and the powers $(p_x,p_y)$ depend on
the initial action. Exponents less than 1 indicate sub-diffusive 
behavior while exponents greater than 1 imply super-diffusive motion
Figure \ref{fig: varianceJx} also shows the fits with
this power law. Growth of the variance in the vertical action can also be
fit by a single power law with small values of $(C_y,p_y)$ showing that
there is no appreciable diffusion in that plane. 
On the resonance $2(3 \nu_x - 2 \nu_s) = 2$, there is significant growth  in
the action at amplitudes of 2 and 2.5$\sg$ compared to neighboring
actions both lower and higher. The exception to the single power law fit
occurs at $x=1\sg$ where the variance stays nearly constant after the 
initial transient and then after about 400,000 turns grows by an order of
magnitude over the next 600,000 turns but 
with oscillations in the variance. These oscillations occur 
because of the large sensitivity to the initial angle at this amplitude.
The oscillations decrease significantly when the number of particles at the 
same initial action is increased from 4000 to 20000 particles, which results in
a more complete sampling of the initial angle. Simulations show that this 
greater sensitivity to the initial angle is also present at amplitudes
in the range $1.0 \sg \le x \le 1.3 \sg$ with $y=0.1\sg$.
The fits to a power law in this zone are applied after the variance starts to 
grow rapidly but with 20000 particles. The average action with initial
$|x|=1.0\sg$ grows about 10\% after 10$^6$ turns while the average action with
initial $|x|=1.5\sg$ grows by about a factor of two over this time. So the
narrow zone around $|x|=1.0\sg$ corresponds to a zone of bounded chaos. 

 At the resonance $4\nu_x - 2 \nu_s = 1$, the growth in variance is largest 
in the range $x=2.5 - 3 \sg$ and drops for both smaller and larger initial
actions. 
The large oscillations in the variance occur in a range around $x=2.0\sg$ and
again these oscillations are reduced when the number of particles is
increased from 4000 to 20,000. For this resonance, the zone around
$|x|=2.0\sg$  is a zone of
bounded chaos. Similar behaviour is seen at other values of $y$ but the
width of the zone of bounded chaos changes.

\begin{figure}
\centering
\includegraphics[scale=0.25,angle=-90]{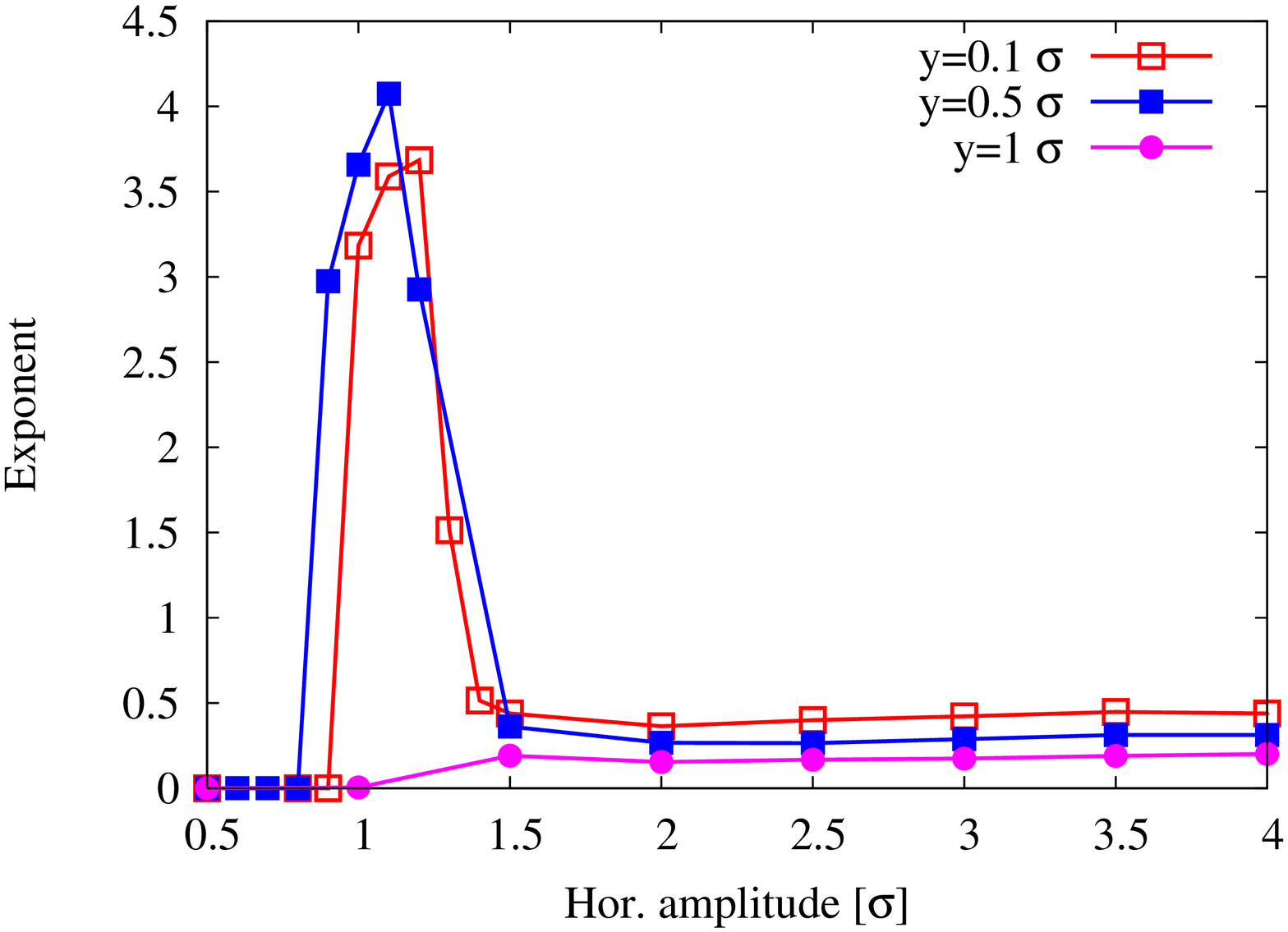}
\includegraphics[scale=0.25,angle=-90]{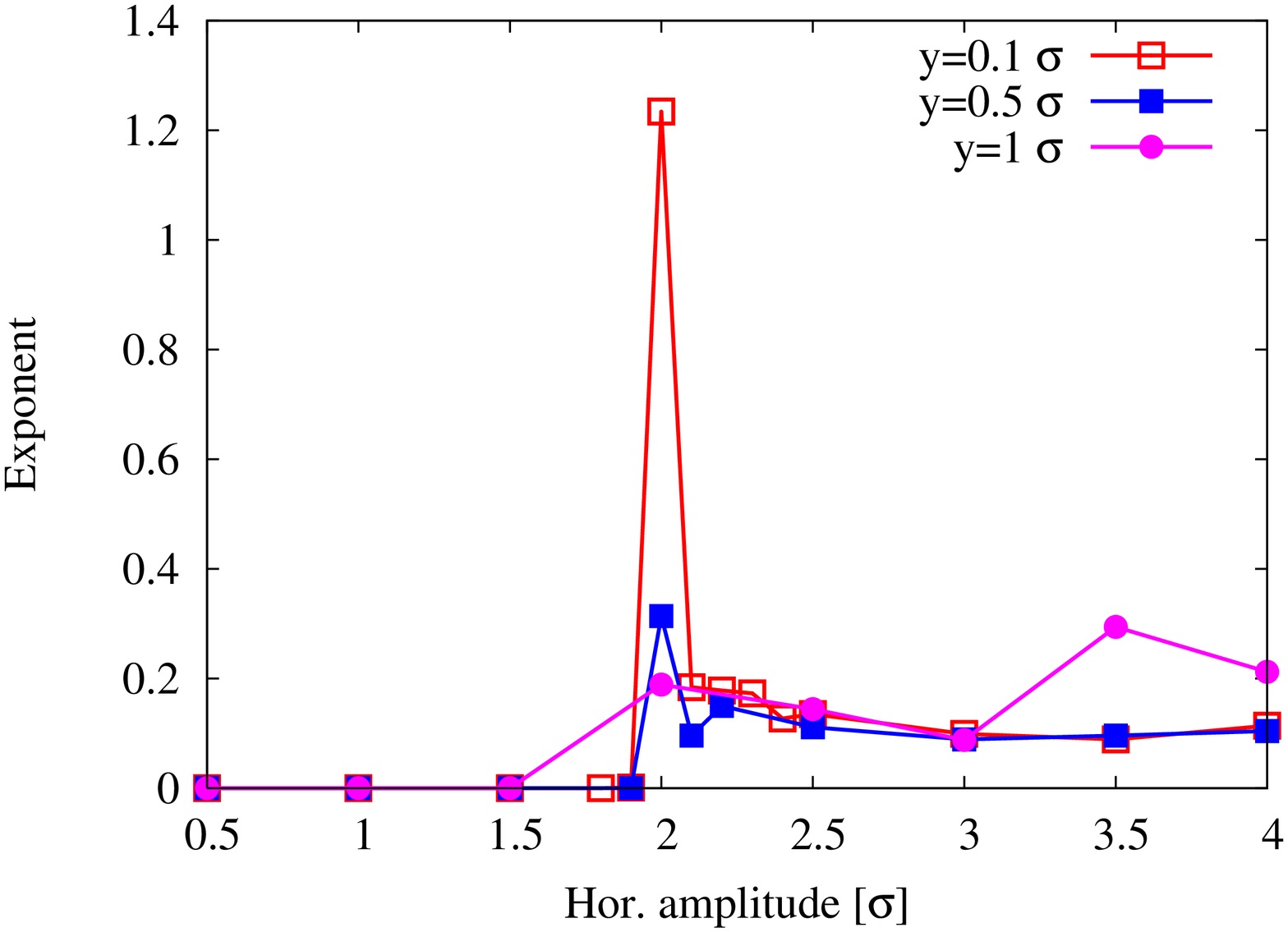}
\caption{(color) Exponent $p_x$ of time in the power law fits of the horizontal 
action variance
vs the initial horizontal amplitude for different initial values of the
vertical amplitude. The left figure corresponds to the 
resonances $2(3 \nu_x - 2 \nu_s) = 2$ and the right to the resonance
 $4\nu_x - 2 \nu_s = 1$. The exponent spikes above 1 in a very narrow 
range of horizontal amplitudes. Exponent values below 1 indicate
 sub-diffusive behavior while those above indicate super-diffusive
behaviour.}
\label{fig: powerfits_Jx}
\end{figure}

The exponents in the power laws were calculated for several values of the 
horizontal amplitude and for different vertical initial amplitudes.
Fit \ref{fig: powerfits_Jx} shows the exponents for both resonances. 
On the resonance $2(3 \nu_x - 2 \nu_s) = 2$ there is a spike in the exponent
to values well above 1 in the regions of bounded chaos for $y=0.1, 0.5 \sg$
suggesting super-diffusive behavior. Above the zone of bounded chaos, the
exponent falls well below 1 suggesting sub-diffusive behavior. At $y=1\sg$
the exponent stays well below 1 for all $x$ showing that zones of bounded
chaos have disappeared. On the  $4\nu_x - 2 \nu_s = 1$ resonance, the exponent
rises above 1only in a narrow zone around $x=2\sg$ at $y=0.1\sg$. At $y=0.5\sg$
the exponents stay well below 1 at all $x$ with a small spike at $x=2\sg$.
The motion is sub-diffusive at all $x$  values studied when $y=1\sg$.
Since the super-diffusive regions are narrow, it is possible that they may
appear for $|y| \ge 1\sg$ when the motion is studied with a finer resolution or 
even when the longitudinal variables are changed.
We remark that we have observed here three different signatures of bounded 
chaos: large variations in final amplitude when starting from the same 
initial amplitude (seen in Fig \ref{fig: ampx_distribs}), 
large oscillations in the action variance over time (seen in Fig.
\ref{fig: varianceJx} and a spike in the power law for the growth of the 
variance (seen in Fig. \ref{fig: powerfits_Jx}). These signatures
apply to an ensemble of particles at the same amplitude but different
initial angle as opposed to the Lyapunov exponent
criterion which is applied to a pair of particles that are initially 
infinitesimally close.

\begin{figure}
\centering
\includegraphics[scale=0.5]{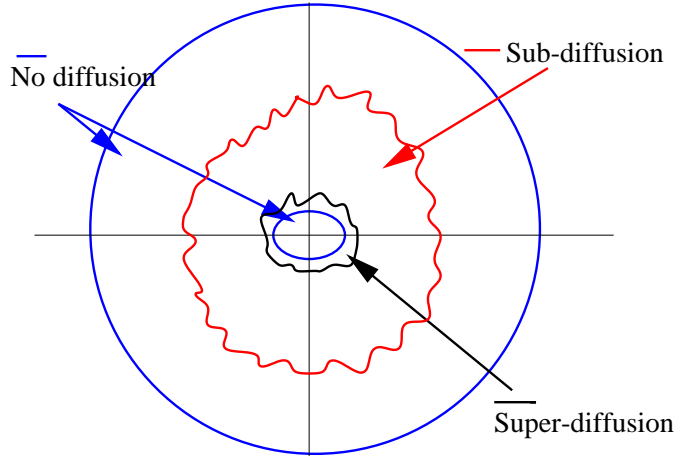}
\caption{(color) Qualitative sketch of phase space divided into zones of no diffusion,
super-diffusion and sub-diffusion when the dynamics is dominated by a 
beam-beam synchro-betatron resonance.}
\label{fig: phasespace}
\end{figure}

The picture that emerges is that near synchro-betatron resonances, phase space 
is divided into
several zones. At small amplitudes there is no diffusion.
At larger amplitudes there is a zone
of bounded chaos with super-diffusive motion. The next zone outward in 
phase space is wider with sub-diffusive motion. Finally at even larger
amplitudes, the motion becomes linear again and consequently there is no diffusion.
Fig. \ref{fig: phasespace} shows a qualitative sketch of these different zones.
The width of the super-diffusive zone with bounded chaos depends on the resonance, on
the amplitude of the orthogonal transverse amplitude and on the values
of the longitudinal variables.  

The fact that the sub-diffusive regions seem to be dominant in this 
perturbed Hamiltonian system should not be unexpected due to the existence of
hyperbolic fixed points and the existence of  perturbed
 KAM tori. These fixed points and tori lead to orbits which stay in their 
vicinity for long 
time periods and consequently to slower growth. Similar phenomena have been
reported for the standard map by Balescu \cite{Balescu}.

\section{Statistics of single particle behavior} \label{sec: single}

We saw in the previous section that in most regions of phase space, the variance
grows slower than linearly with time. If we define an instantaneous or
'running' diffusion
coefficient \cite{Balescu} as $D_{J_x} = (1/2)\del \lan \Dl J_x^2 \ran/\del t$, then this 
coefficient would be time dependent and would vanish in the very long time limit.
Near both resonances we did not observe any zone of regular diffusion with
constant diffusion coefficients. 
We also saw that the beam profile was given by a Levy stable distribution
which is known to be the solution of a fractional diffusion equation. 
These suggest that the dynamics near these resonances cannot be described
by the regular diffusion equation but instead that the diffusion is anomalous
which needs a different diffusion equation. In order to test this possibility
in more detail, we will examine the validity of the assumptions behind the 
regular diffusion equation.

\subsection{Continuous Time Random Walks}

The regular diffusion equation arises after assuming that the particle
dynamics can be modeled as a classical random walk following a Markov process. 
This implies that particle jumps occur at regular time intervals and there is a
well defined time scale such that events separated in time by longer than this time
scale are uncorrelated. It then follows that the particle density is governed
by the well known Chapman-Kolmogorov master equation. From this master 
equation and a few more assumptions (e.g. on the smallness of the 
displacements etc.) the regular diffusion equation follows. See Appendix A
for a sketch of this derivation. 

A well known alternative to the standard random walk picture is the 
Continuous Time Random Walk (CTRW) model introduced by
Montroll and Weiss \cite{MontrollWeiss}  to consider processes where
both the times at which jumps occur as well as the sizes of the jumps in
space are random functions. A review of CTRW and connections to fractional
diffusion equations may be found in \cite{Metzler}.

A general dynamical process may not have a characteristic time scale. 
In those cases a Markov description may not be applicable. 
The CTRW model introduces the concepts of 
a probability distribution $w$ for the waiting times before a jump occurs and
a probability distribution $\Psi$ for the size of a jump 
In beam dynamics there is no diffusion when the motion is linear and the
usual Courant-Snyder actions are conserved. Consequently it makes sense to
define the jumps in action space when the motion is nonlinear and diffusive. 
Hence we define $w(t,{\bf J}) \Dl t $ to be the probability that a particle waits
for a time between $t$ and $t+ \Dl t$ at action ${\bf J}$ before making a jump.
and define $\Psi(\Dl {\bf J};{\bf J},t)\Dl {\bf J} $ to be the probability of 
making a jump by $\Dl {\bf J}$ at the action ${\bf J}$ at time t. 
These distributions are normalized, i.e.
\beq
\int w(t,{\bf J}) dt = 1 = \int \Psi({\bf J'};{\bf J},t) d{\bf J'}
\eeq
The concept of a waiting time endows the system with memory. 
The CTRW model reduces to the classical random
walk model on which the regular diffusion equation is based, when
the waiting time follows an exponential behavior in time $e^{-t/\tau}$ with
a characteristic time scale $\tau$. 

These waiting time and jump size distributions can be used in many cases to 
determine the evolution followed
by the density distribution $\rho({\bf J},t)$. The canonical 
CTRW model assumes a power law waiting time distribution, a Gaussian
for the jump size distribution and a constant diffusion coefficient.
These lead to a fractional diffusion equation for the density
\cite{Metzler}. 
In our case the dynamics near resonances is sufficiently complicated
that we need to establish the evolution equation for the density from
first principles. 
We therefore need to determine the forms of the jump size
distribution and the waiting time distributions from the dynamics.
Simulations discussed in the rest of this section are used to extract
these distributions.

\begin{figure}
\centering  
\includegraphics[scale=0.35]{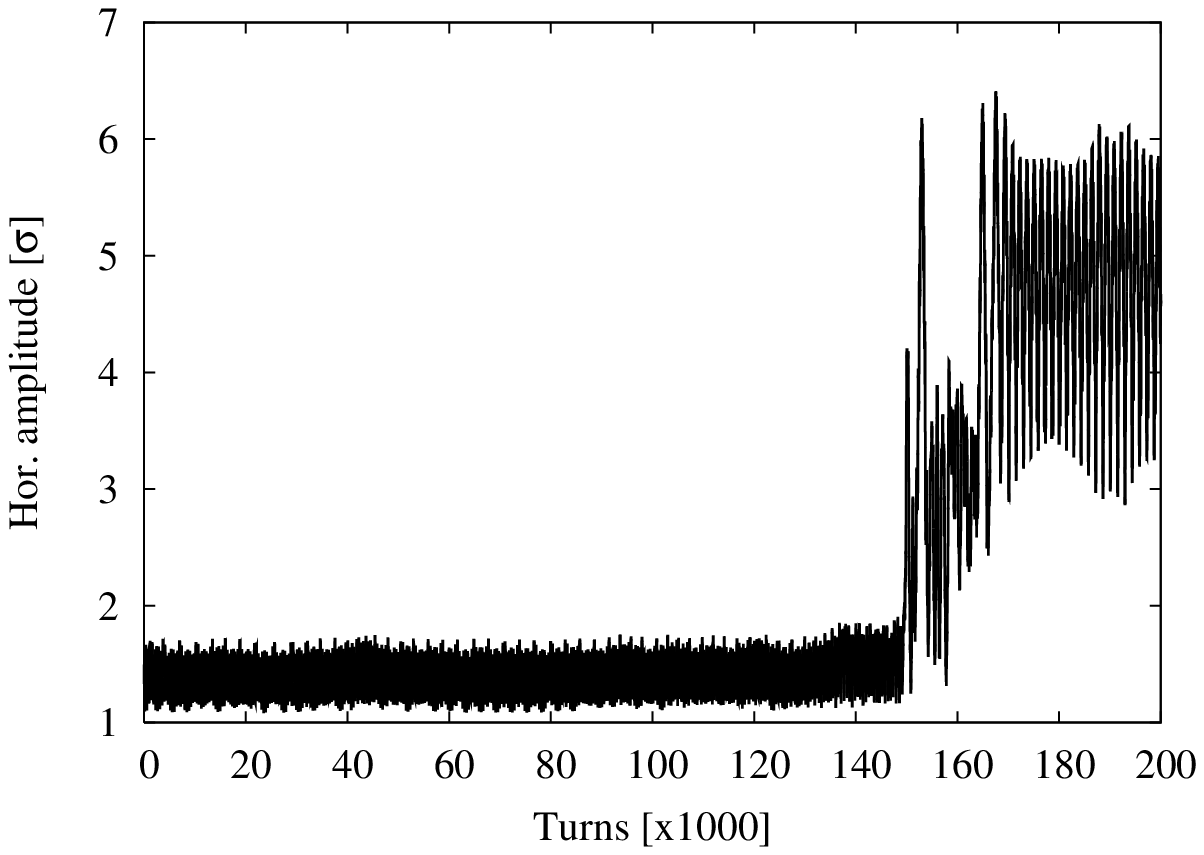}  
\includegraphics[scale=0.35]{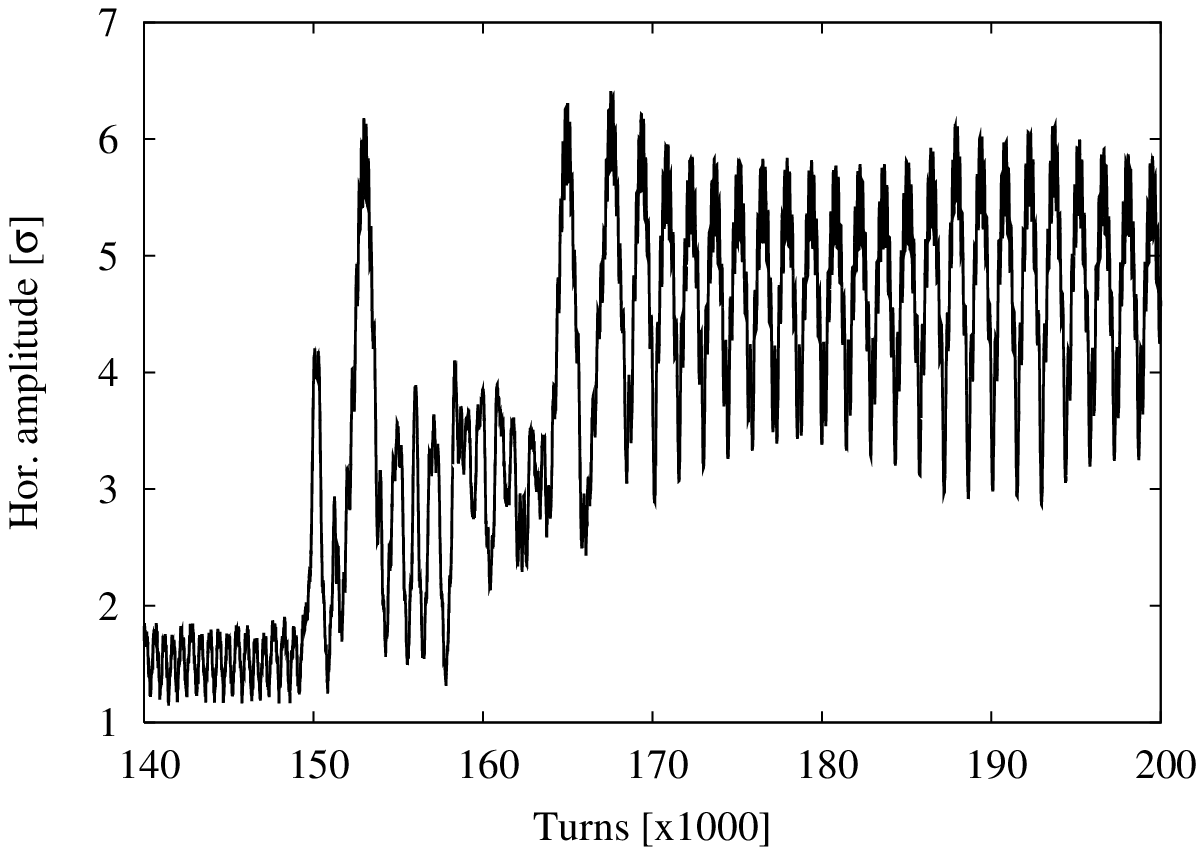} 
\includegraphics[scale=0.35]{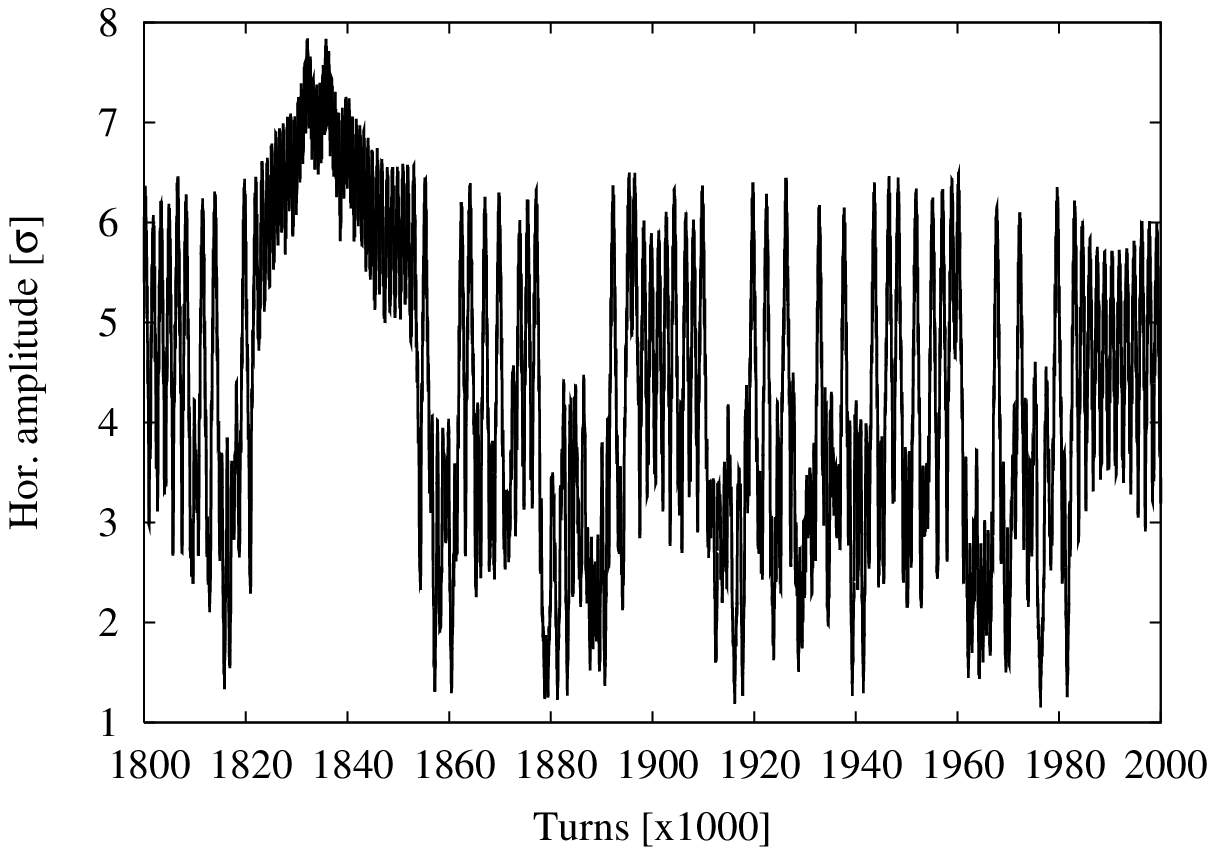}  
\caption{Time series of the horizontal amplitude with $x=1.5 \sg, y=1.5\sg$.
Left plot: We see the first large jump in the amplitude 
from $\sim 1.5 \sg$ to $\sim 6.5 \sg$ occurs after 140,000 turns.
Middle plot: This zoomed in plot shows sequences of jumps from 
small to large amplitudes and back interspersed with intermittent 
periods of small amplitude changes. Right: The last 200,000 turns during an
evolution over 2 $\times 10^6$ turns. Here we see excursions between 
1 to 8 $\sg$.}
\label{fig: Ampx_tseries_1} 
\end{figure}

A check of the whether the CTRW model may be applicable here can be done by
examining the time series of single particles. Fig \ref{fig: Ampx_tseries_1} 
shows one example of a time series of the amplitude $\sqrt{2\bt_x J_x}$ for a
single particle on the resonance $2(3\nu_x=2\nu_s)=2$. 
The left plot shows that a particle may perform small amplitude
quasi-periodic oscillations for a while before a major qualitative change
occurs. The middle and the right plots show that step sizes can be large 
(several $\sg$), of varying amplitude, and there
are intermittent sequences of varying duration where there are smaller steps. 
The time dependent behaviour of this sequence and the non-locality of the changes
establish that this is a process with a distribution of waiting times and
a distribution of action step sizes, the key ingredients of the CTRW model.

\subsection{Jump size distributions}

\begin{figure}
\centering
\includegraphics[scale=0.35]{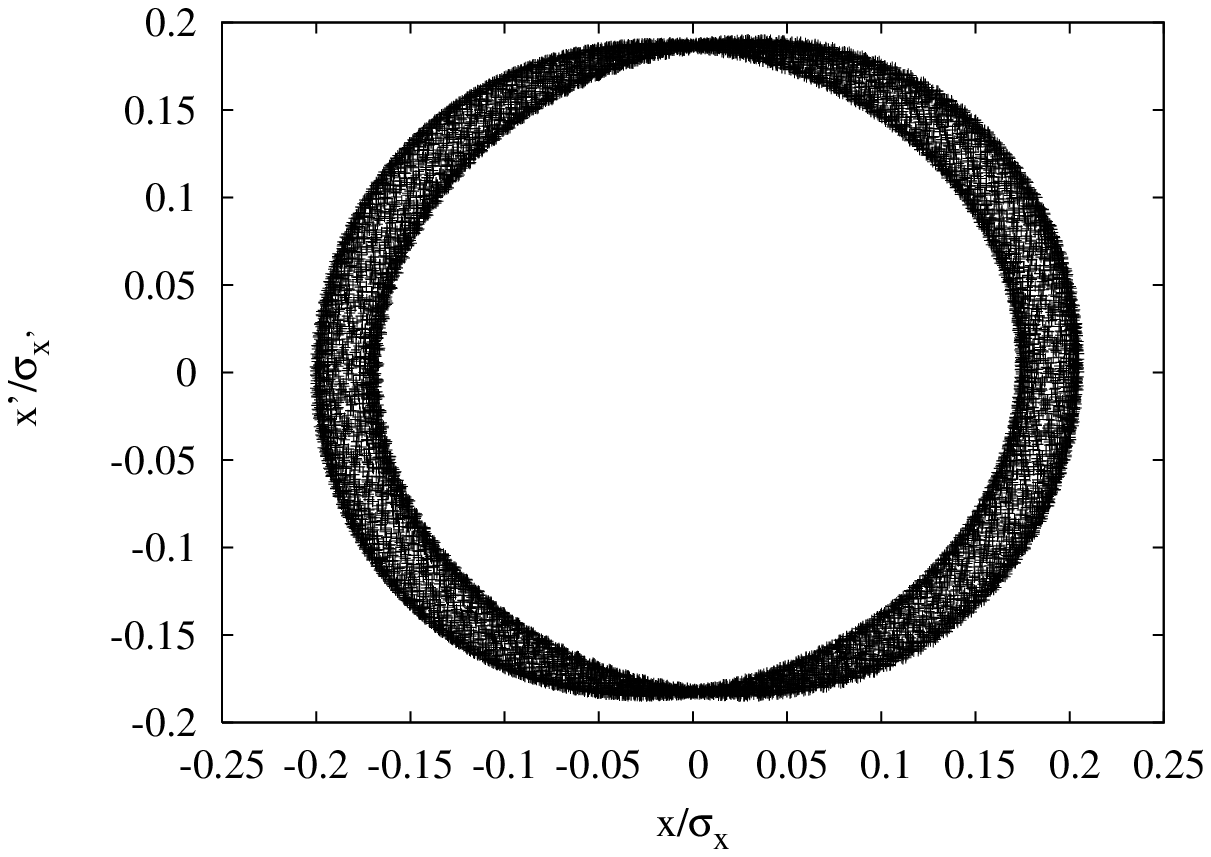} 
\includegraphics[scale=0.35]{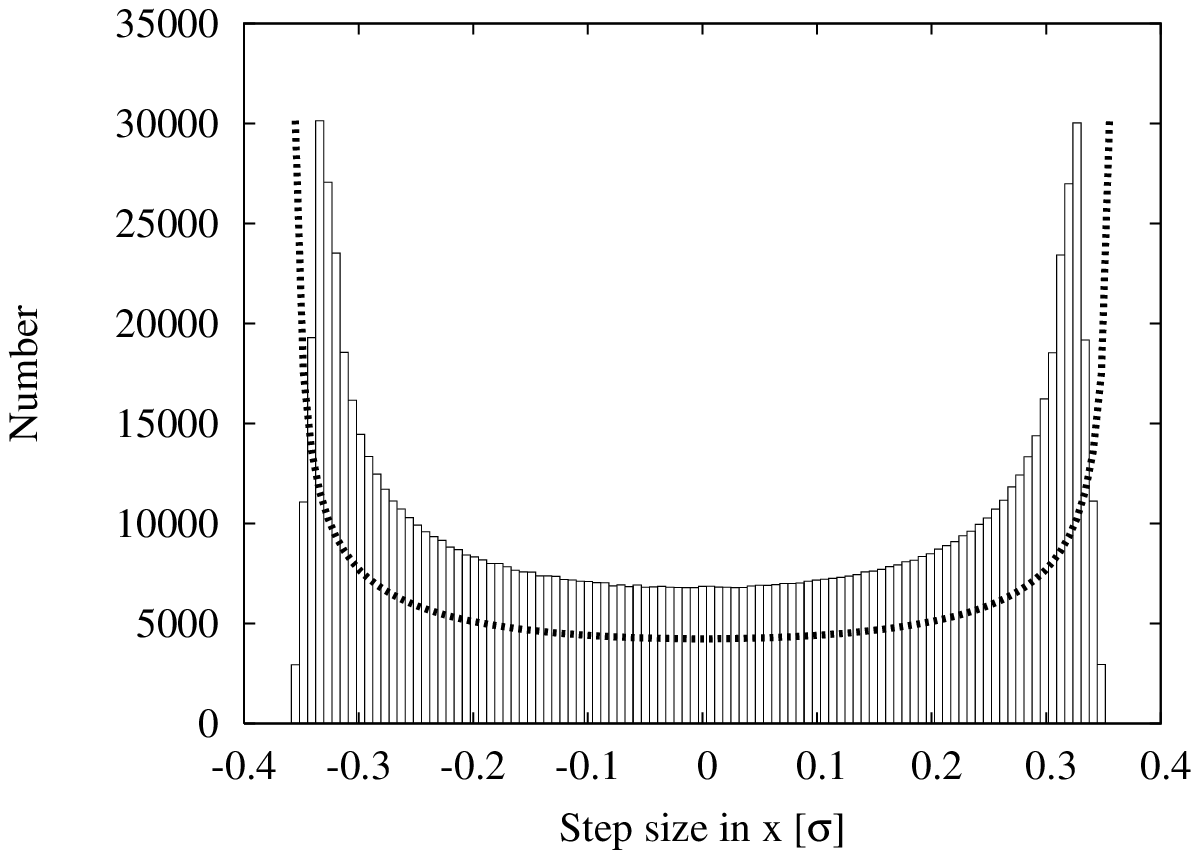} 
\includegraphics[scale=0.35]{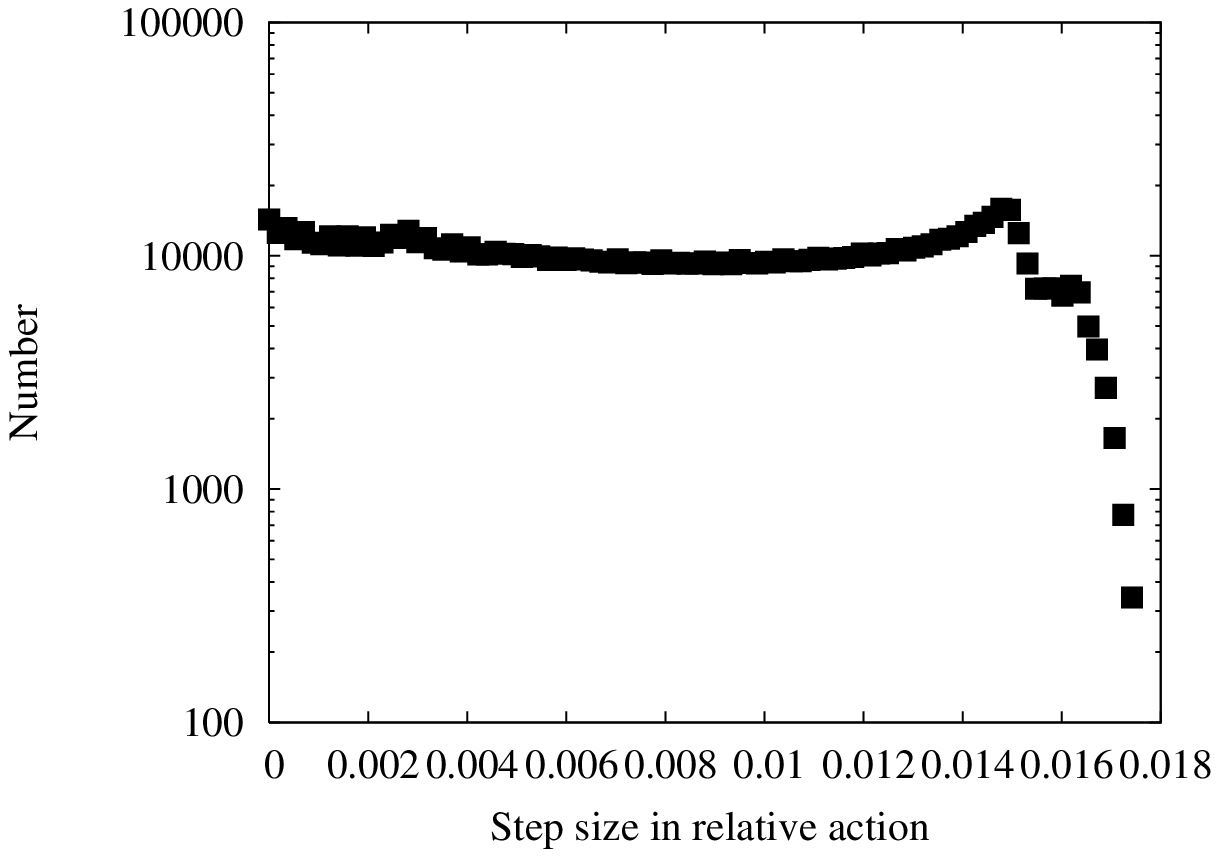} 
\includegraphics[scale=0.35]{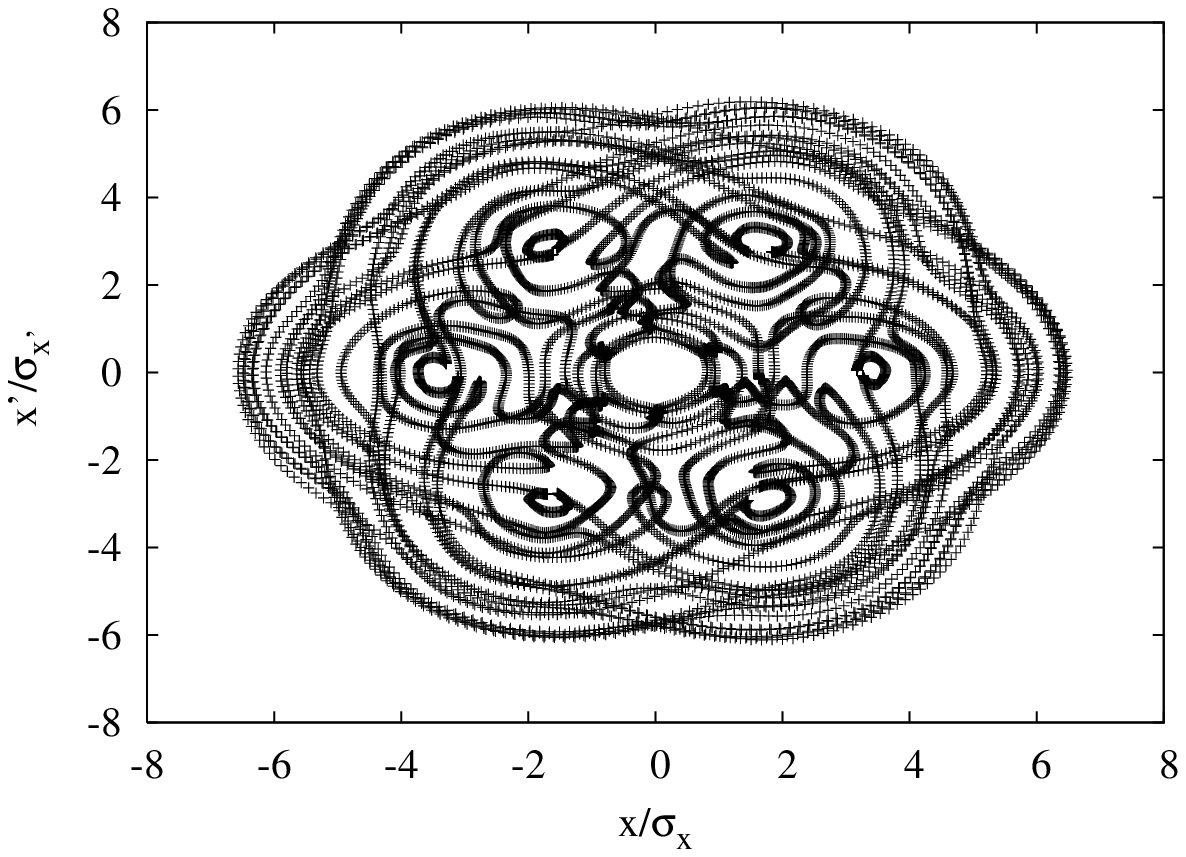} 
\includegraphics[scale=0.35]{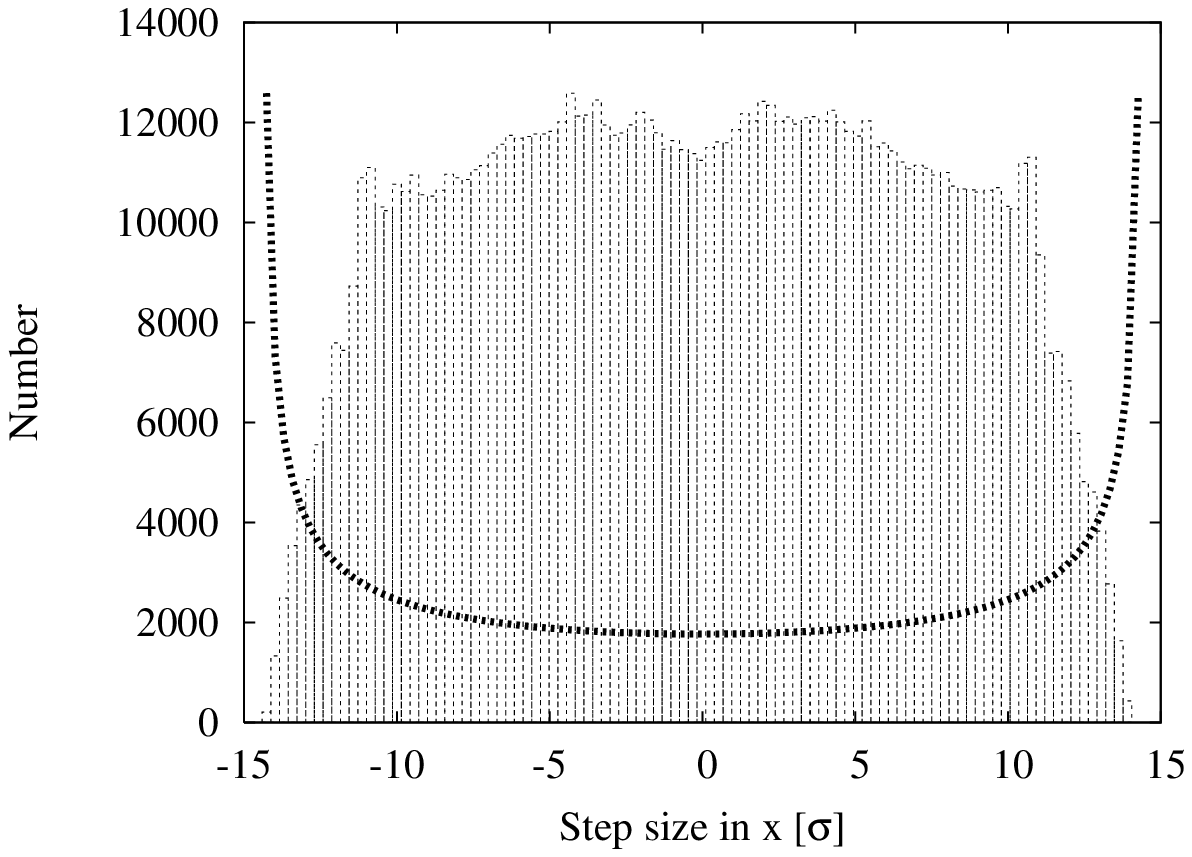} 
\includegraphics[scale=0.35]{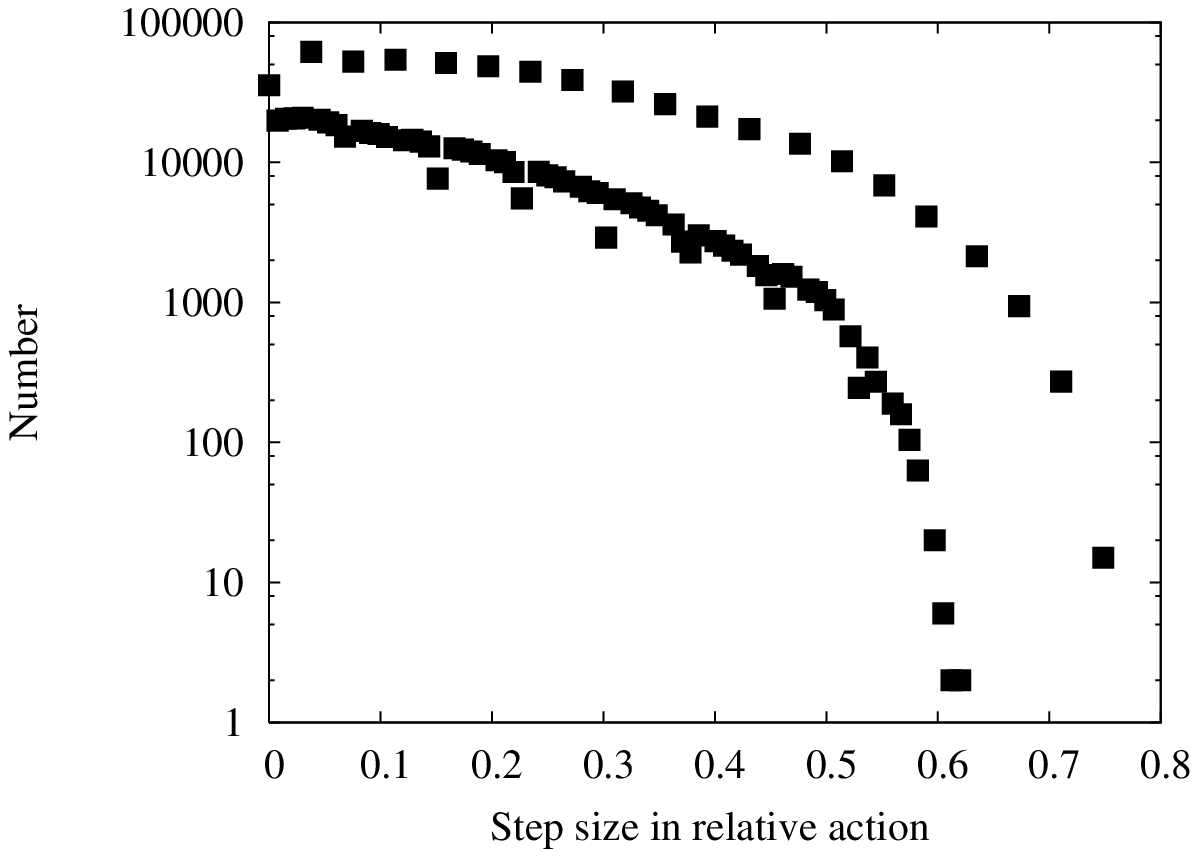} 
\includegraphics[scale=0.35]{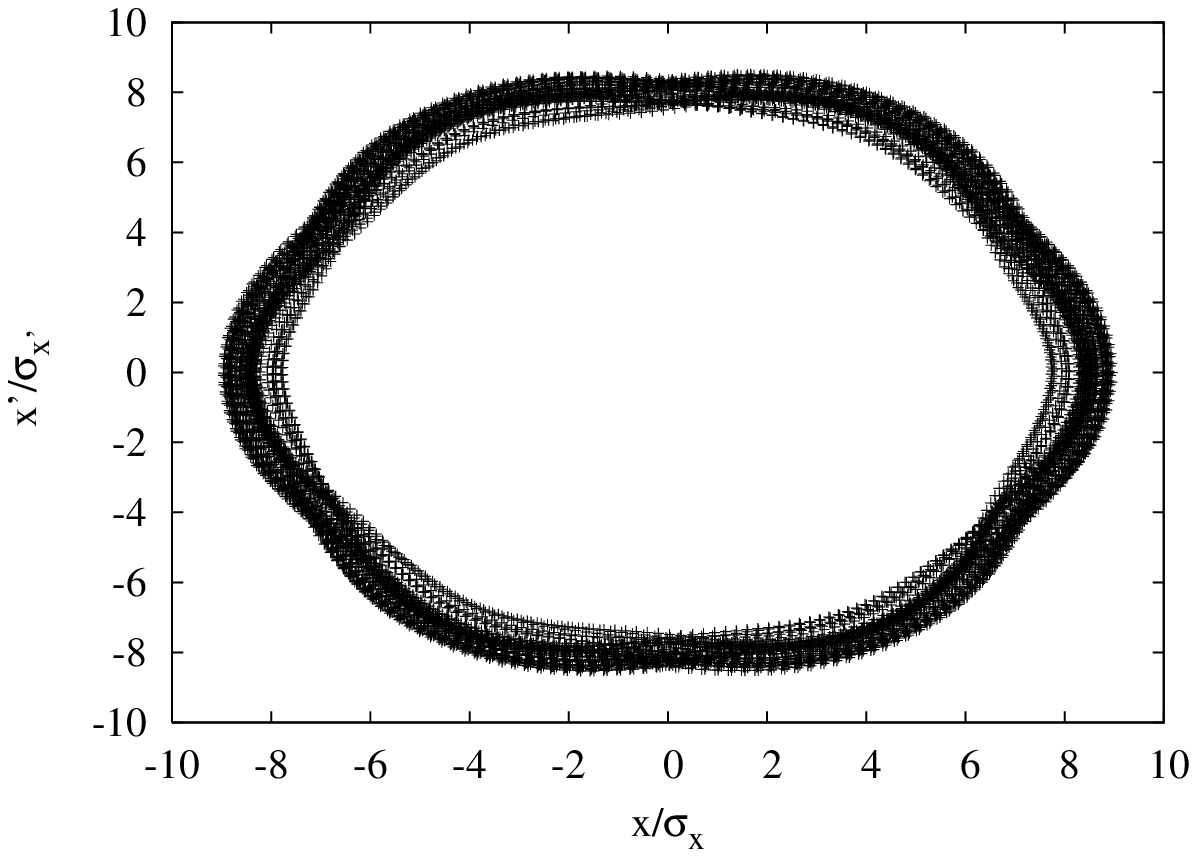} 
\includegraphics[scale=0.35]{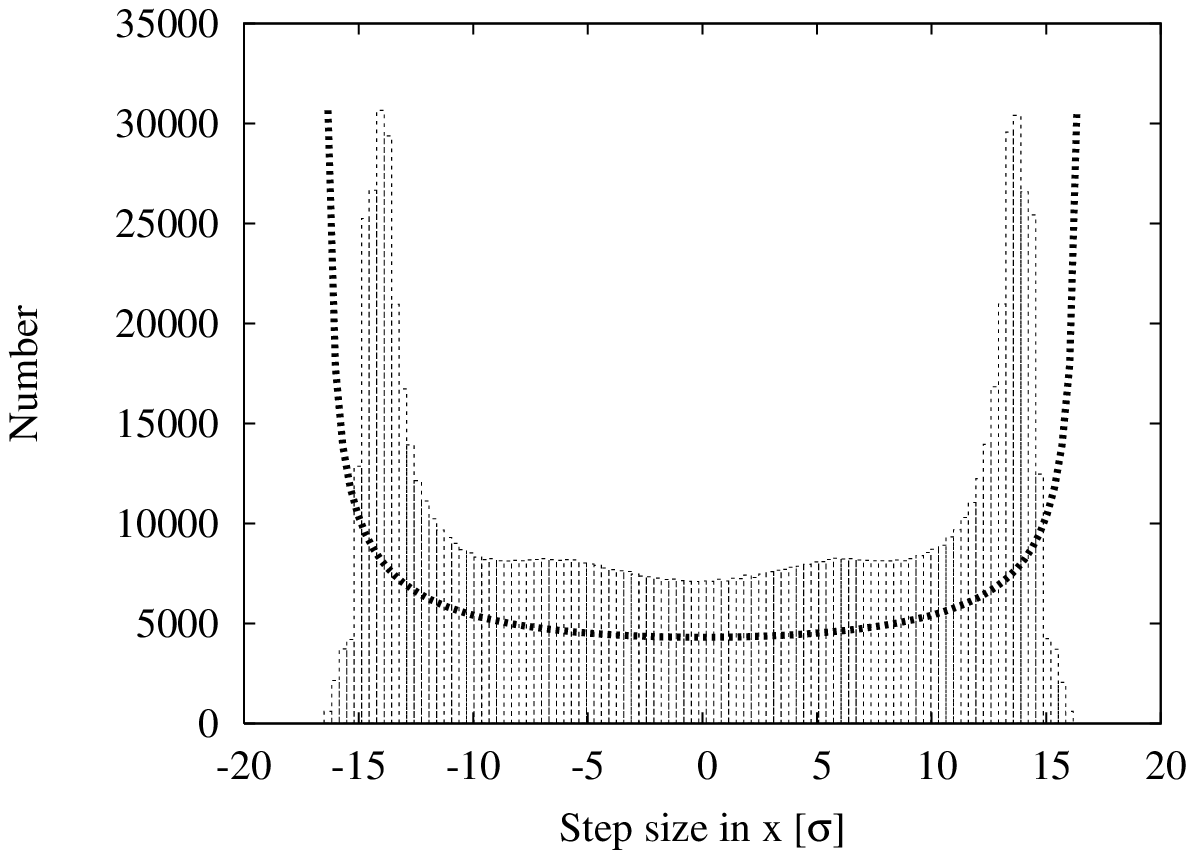} 
\includegraphics[scale=0.35]{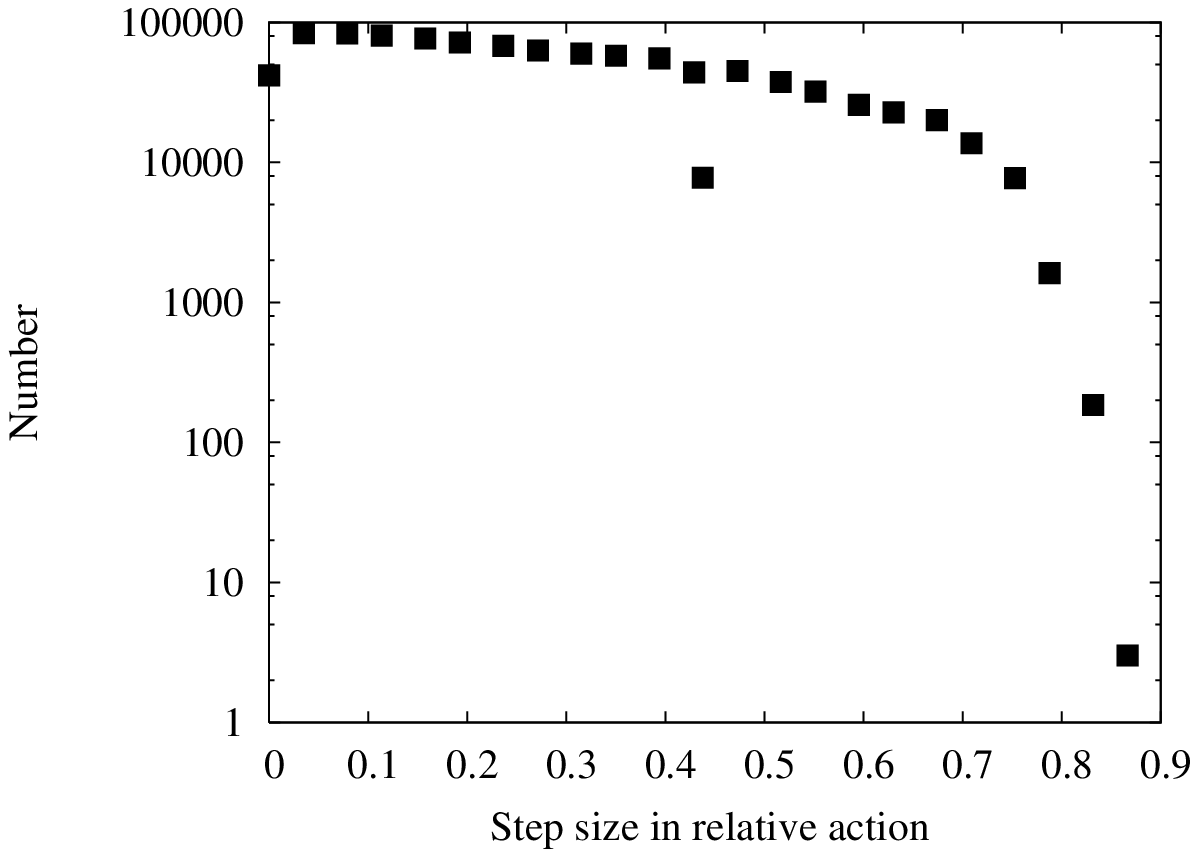} 
\caption{Phase space (left), jump size distribution in $x$ (middle) 
and jump size distribution in action $J_x$ (right) on the resonance 
$2(3 \nu_x - 2 \nu_s) = 2$. The initial value of $x$ changes going 
from top to bottom as $x = (0.2, 2.0, 8.0)\sg$.
The initial value of $y=0.1 \sg$ is the same in all these plots.
The distribution in $\Dl J_x$ is plotted on a semi-log scale and the
abscissa is in units of $\Dl J_x/J_{\sg}$ where $J_{\sg}$ is the action at 1 $\sg_x$.}
\label{fig: xpspace_3rd}
\end{figure}

We now calculate the jump size distributions by following a single
particle for 10$^6$ turns and find the changes $\Dl x, \Dl J_x$
in position and action per turn. Fig. \ref{fig: xpspace_3rd} shows
the phase space, and jump distributions of $\Dl x, \Dl J_x$ on
the resonance $2(3\nu_x-2\nu_s)=2$ with initial values of $x=(0.2, 2, 8)\sg$. 
At the smallest initial 
position $x_0 = 0.2\sg$, the phase space is  a distorted ellipse
with no trace of the resonance island; motion here is quasi-linear.
The plot for the distribution function of $\Dl x$ also has the
distribution function for a periodic function shown in dotted lines.
When the argument of a periodic function like sine or cosine is 
sampled from a random distribution, the distribution function for
the periodic function $f$ has the form
\beq
p(f) \sim \frac{1}{\sqrt{1 - f^2}}, \;\;\;\;\;\;\;\;\;\;\;\; |f| \le 1
\eeq
The distribution function has local maxima wherever the function
itself becomes stationary, so that many more points are sampled
from the neighbourhood of these stationary points. Since the motion
at small and large amplitudes is quasi-periodic in our model, it is
to be expected that the distribution in $\Dl x$ is close to that of
a periodic function. The distribution function for $\Dl J_x$ is
plotted on a semi-log scale and shown as discrete points,  for
greater clarity. At $x_0 = 0.2\sg$,
the distribution for $\Dl J_x$ lies on a single curve but not given
by any simple expression. As the particle's initial position 
increases to 2 $\sg$, the nonlinearity of the beam-beam force
manifests and we see resonance islands in phase space and large
excursions. The distribution function for $\Dl x$ undergoes a
qualitative change to resembling a parabolic curve but with a dip
in the center and with peaks close to the center. The distribution
function for $\Dl J_x$ now falls on two separate curves. 
Similar distributions for $\Dl x, \Dl J_x$ are seen for initial particle
amplitudes in the range $1.5\sg \le |x_0| \le 6.5\sg$.
At $x_0 = 8\sg$, the phase space returns to a distorted ellipse with
considerable smear, and the distribution functions also resemble 
those seen at $x_0=0.2\sg$. 

\begin{figure}
\centering
\includegraphics[scale=0.35]{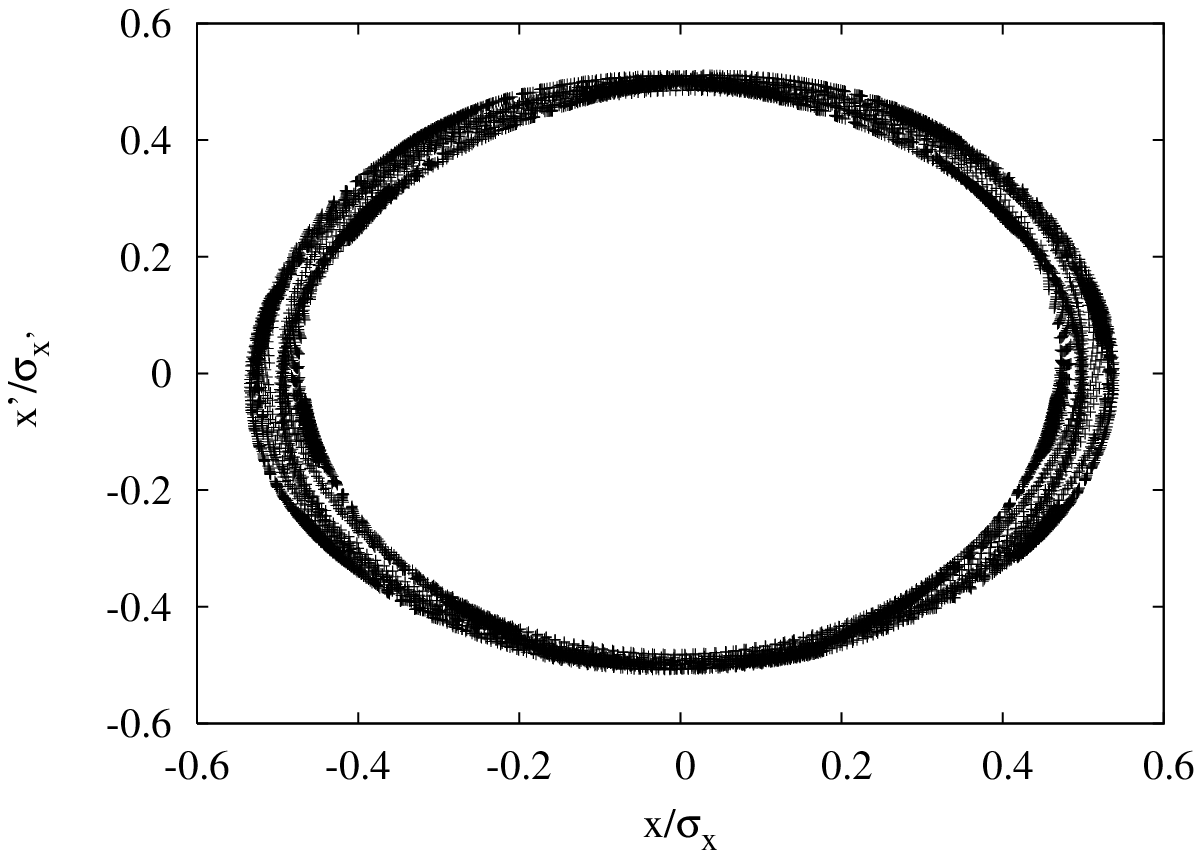} 
\includegraphics[scale=0.35]{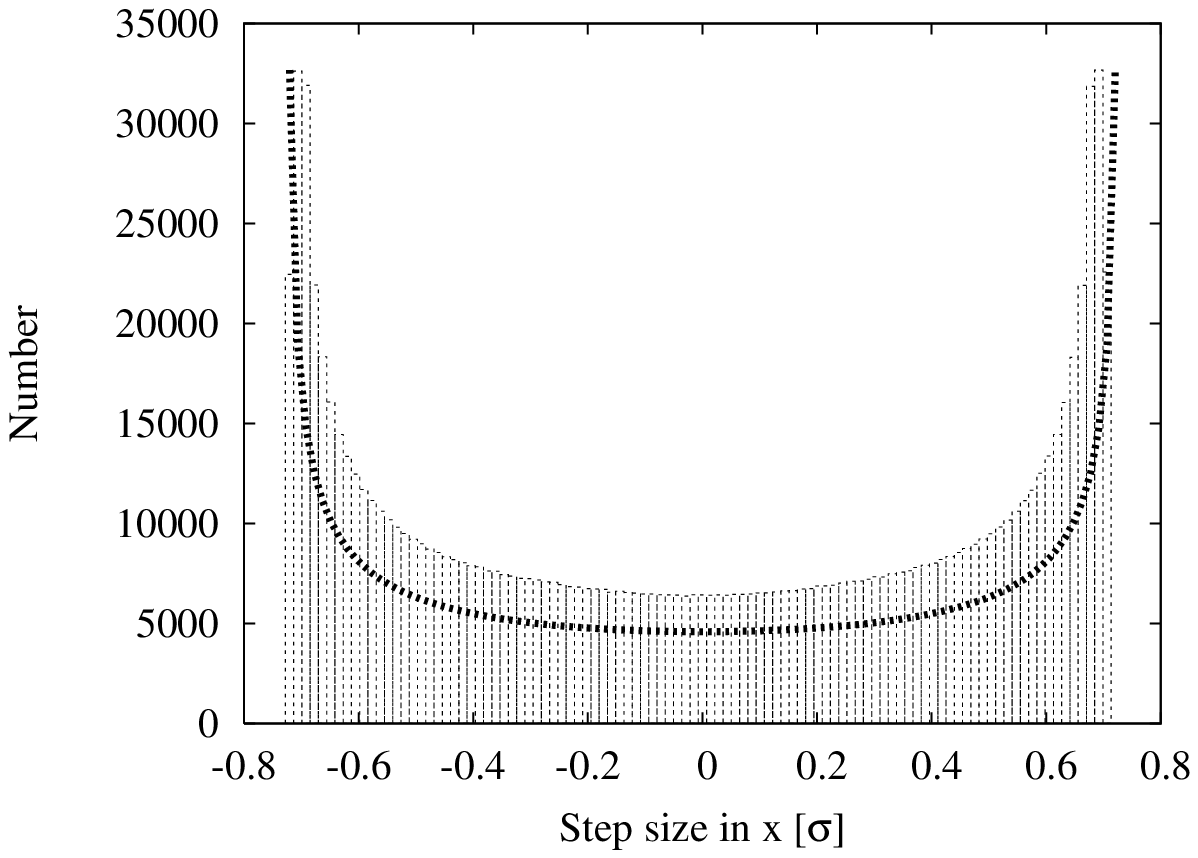} 
\includegraphics[scale=0.35]{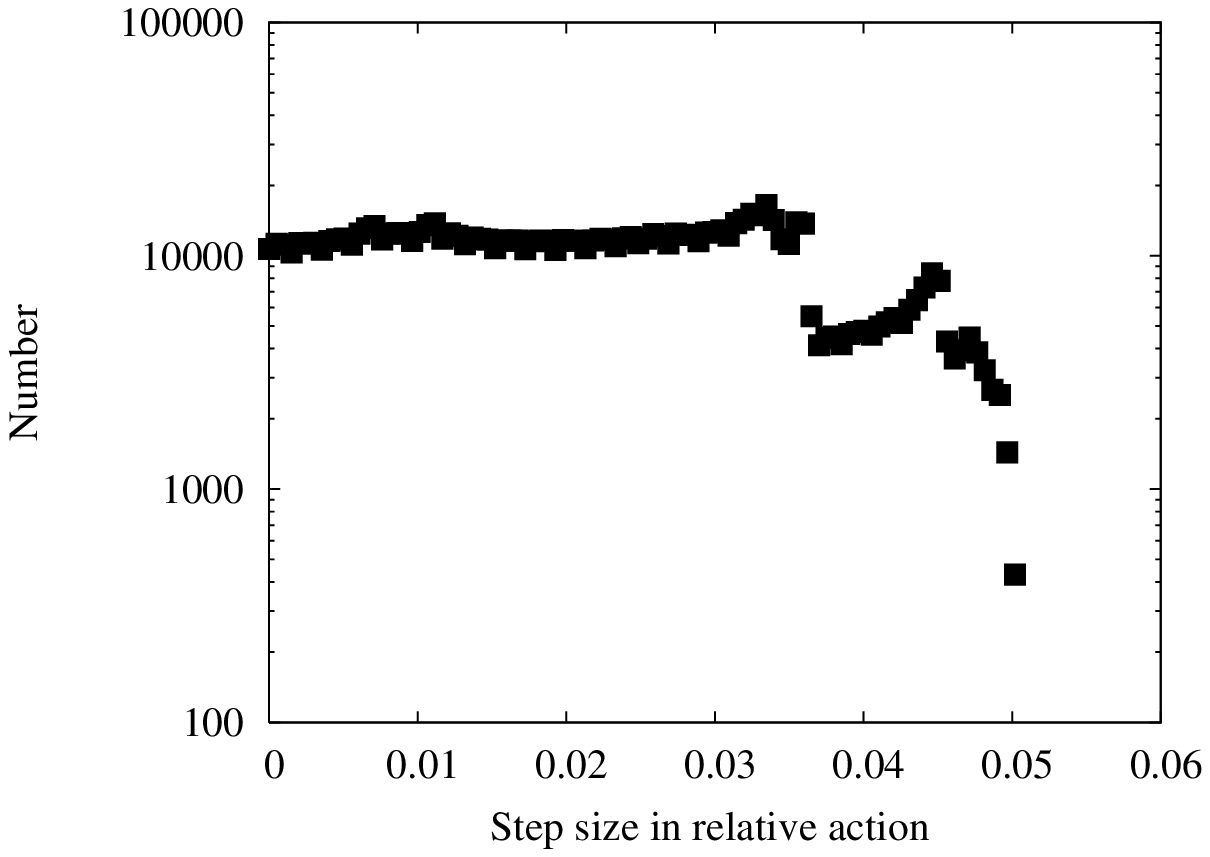} 
\includegraphics[scale=0.35]{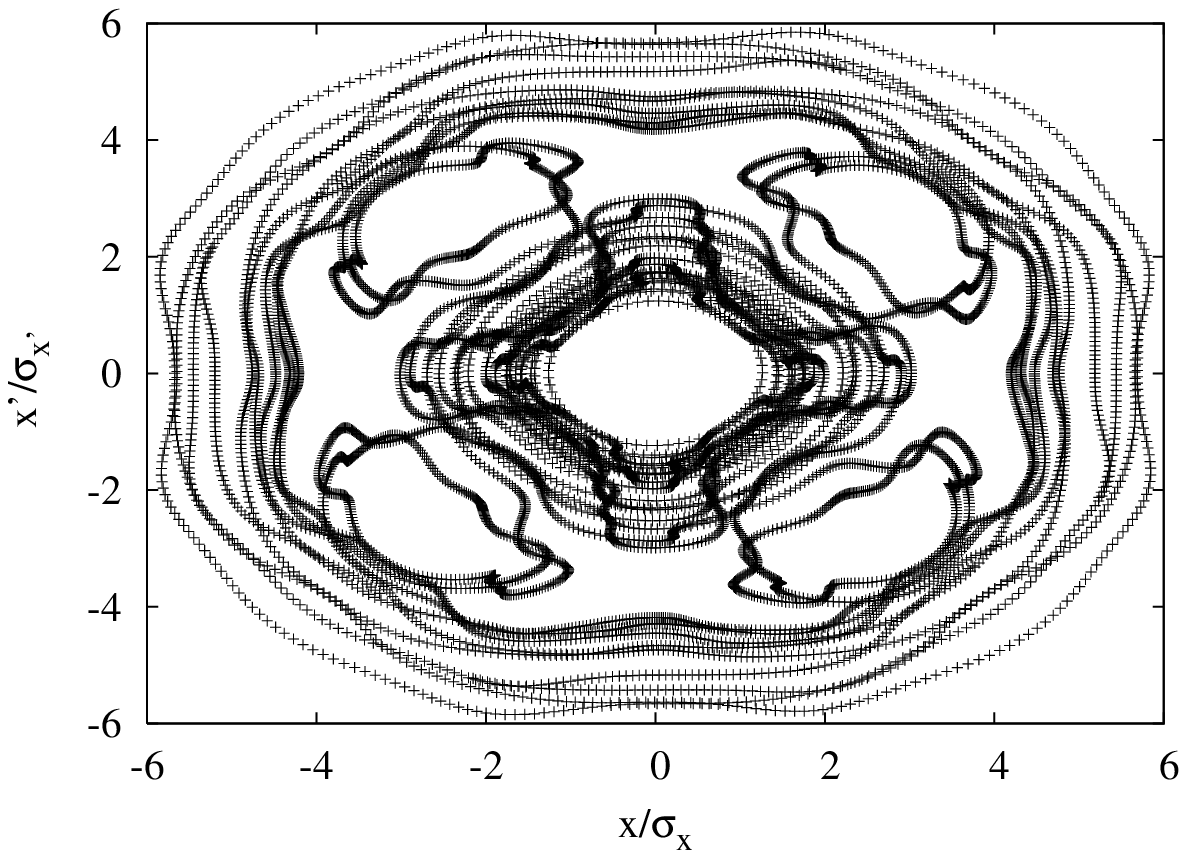} 
\includegraphics[scale=0.35]{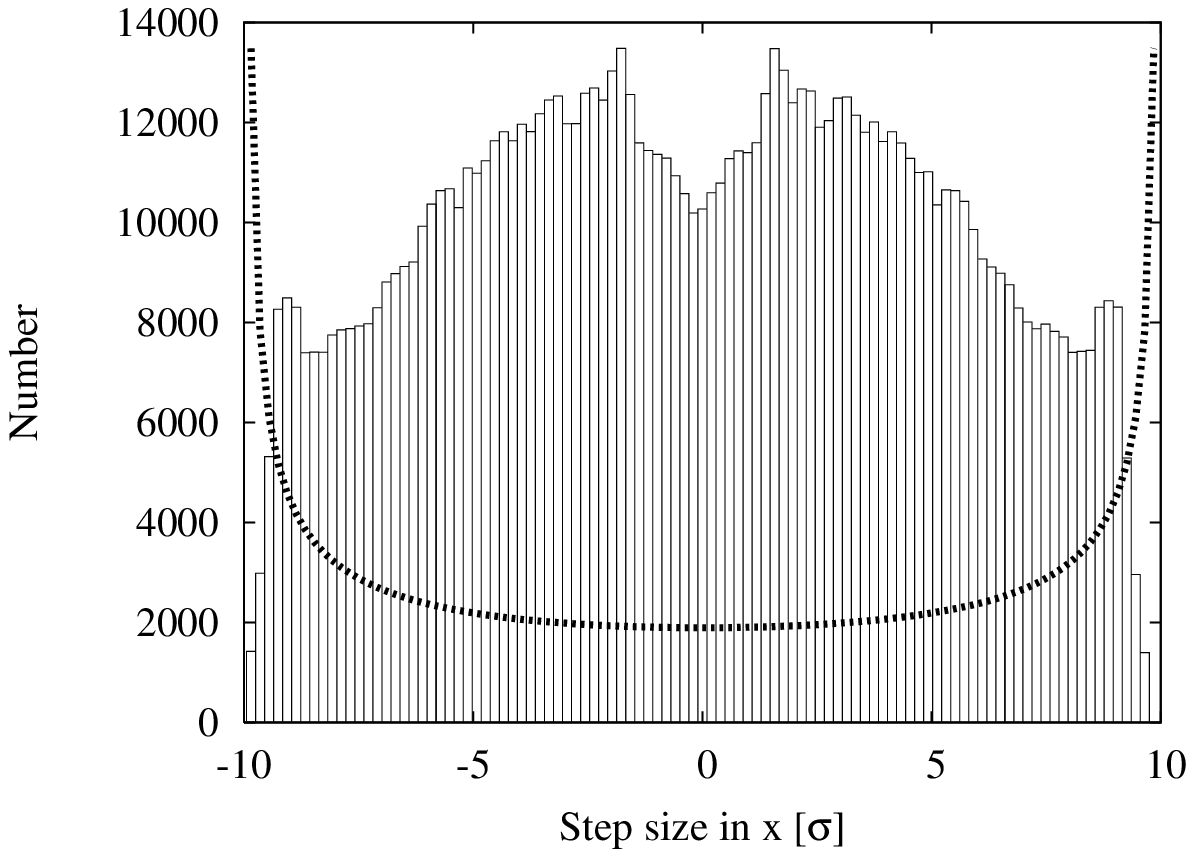} 
\includegraphics[scale=0.35]{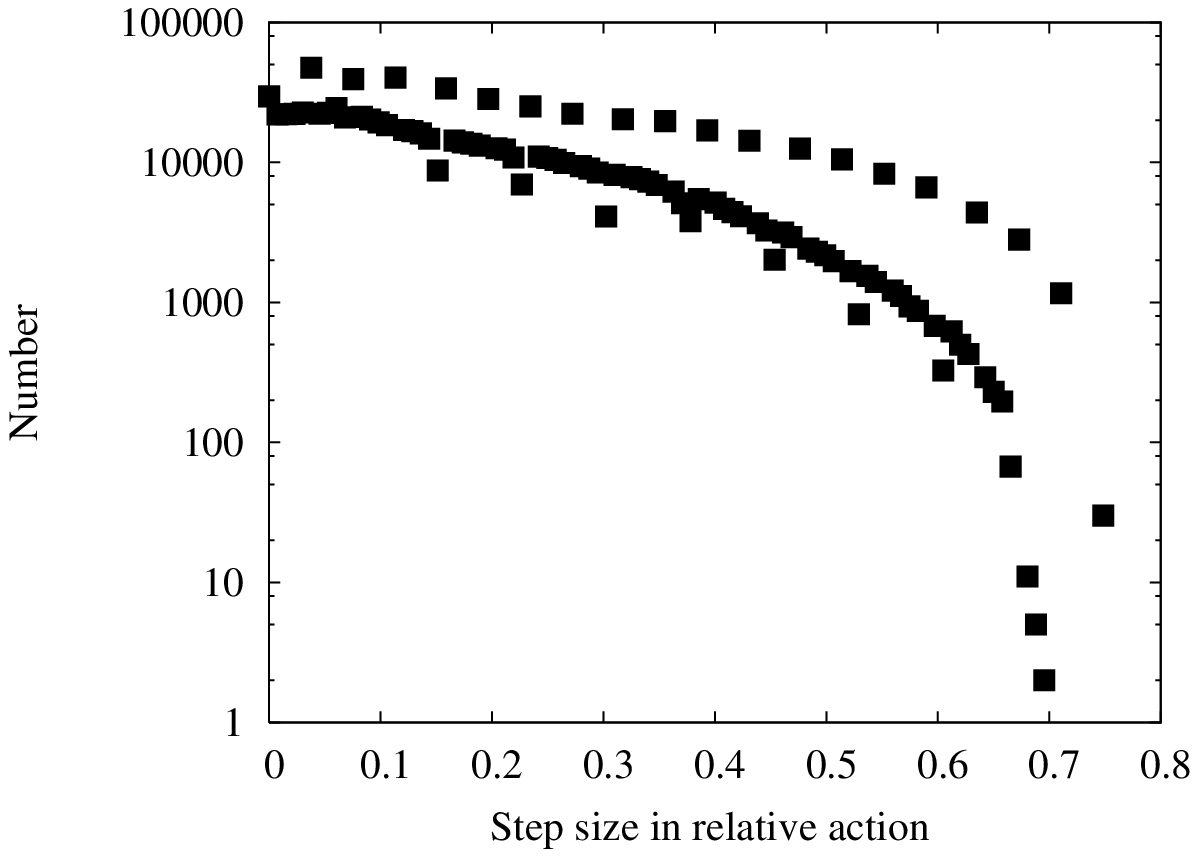} 
\includegraphics[scale=0.35]{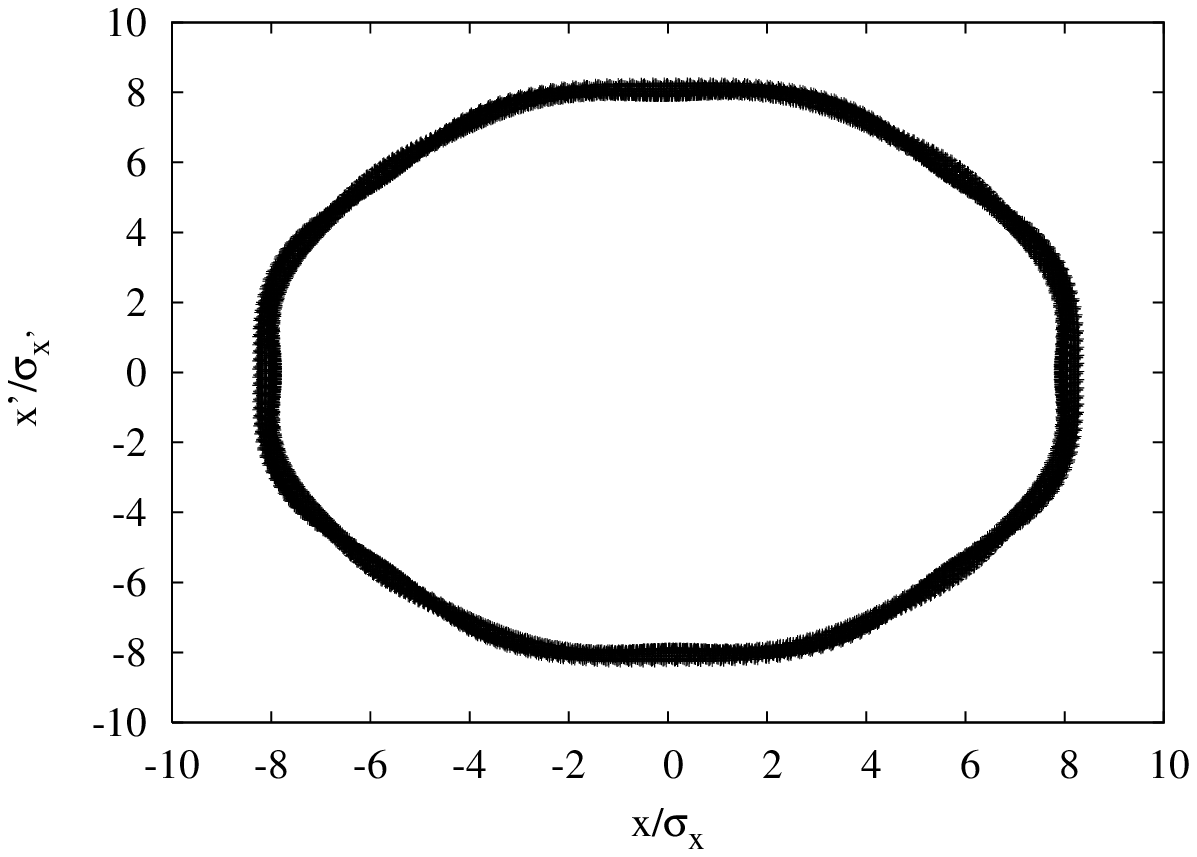} 
\includegraphics[scale=0.35]{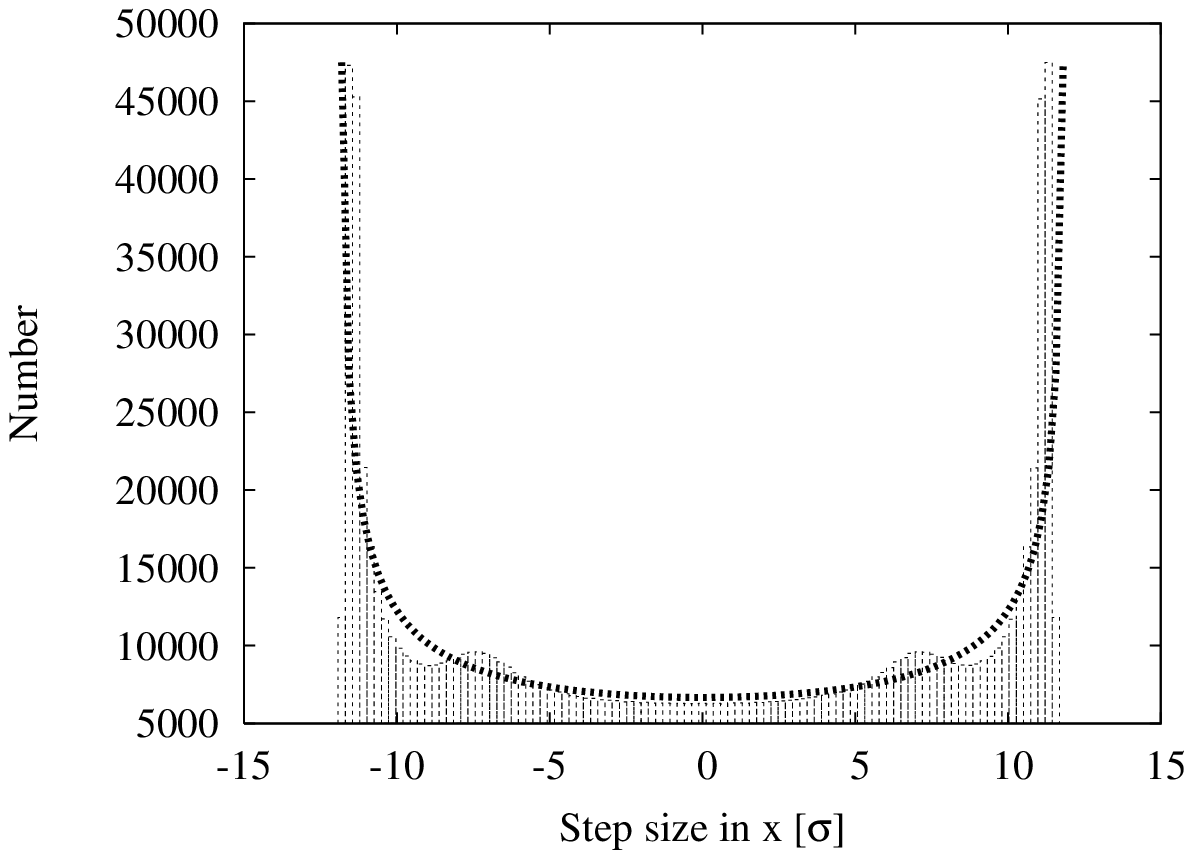} 
\includegraphics[scale=0.35]{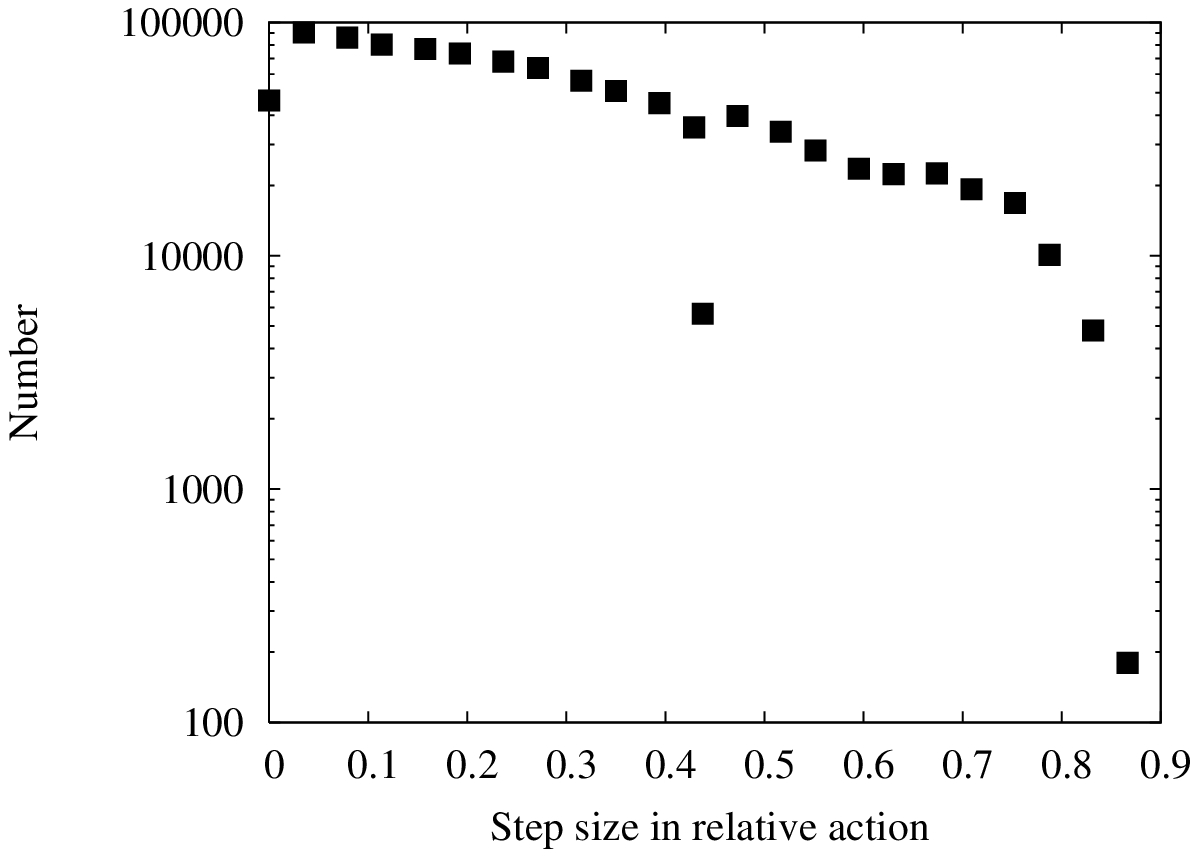} 
\caption{Similar plots as in Fig \ref{fig: xpspace_3rd}
but on the resonance $4\nu_x - 2 \nu_s = 1$.
From top to bottom, the different initial values of $x = (0.5, 5, 8)
\sg$. The initial value of $y = 0.1 \sg$ is the same in all cases. }
\label{fig: xpspace_4th}
\end{figure}
Fig. \ref{fig: xpspace_4th} shows similar plots on the resonance
$4\nu_x-2\nu_s=1$ with initial values of $x=(0.5, 3.5, 8)\sg$.
Again, we see a qualitative change in the distribution functions
when the motion is strongly nonlinear in the presence of the
resonance islands. The shapes of the distributions in $\Dl x$
are similar to those seen for the previous resonance and the
distribution of $\Dl J_x$ also lies on separate curves at 
intermediate amplitudes. These suggest that there is a universal
character to the jump distributions which mirrors the behavior in
phase space.

\subsection{Waiting time distributions}

\begin{figure}
\centering
\includegraphics[scale=0.55]{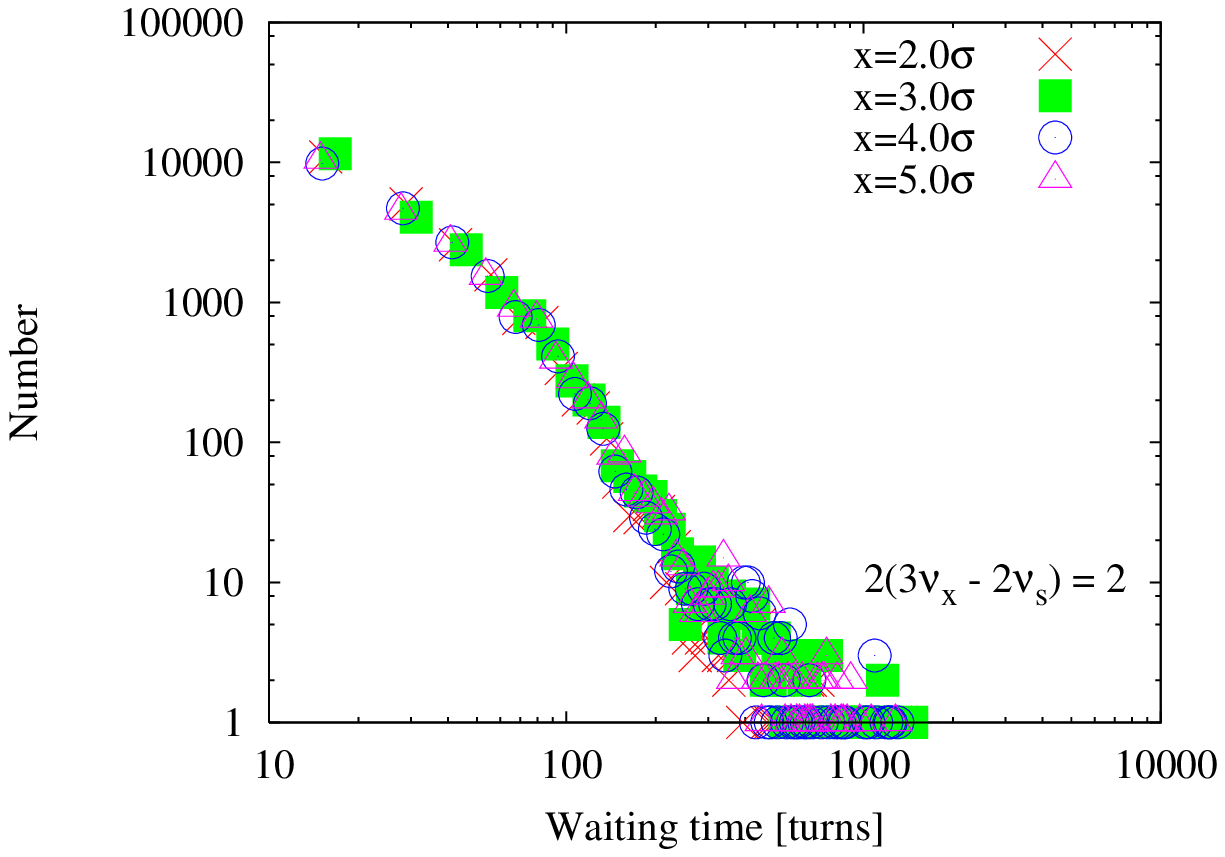} 
\includegraphics[scale=0.55]{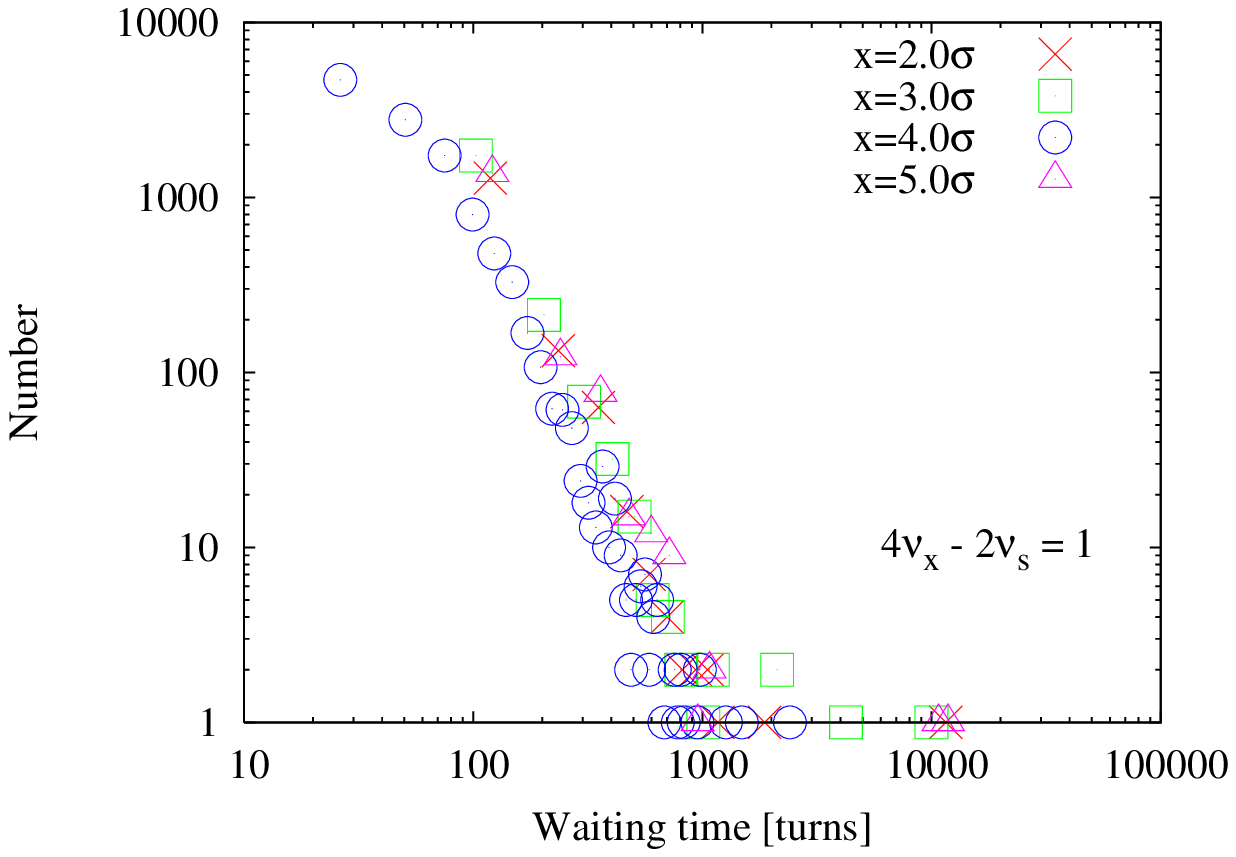} 
\caption{(color) Waiting time distribution (on a log-log scale) for the 
$2(3\nu_x-2\nu_s)=2$ resonance (left) and 
$4\nu_x-2\nu_s=1$ resonance (right). The distributions are calculated
at amplitudes where there is significant diffusion of particles
to larger amplitudes.}
\label{fig: waiting_dist}
\end{figure}

The waiting time distribution is the important distribution that determines
the nature of the diffusion process. As remarked earlier, a waiting time
distribution that follows an exponential law reduces to a Markov process,
otherwise the process is non-Markovian. The waiting time for each initial
amplitude is found here by tracking a particle at that amplitude for 10$^6$
turns. The phase space region in action angle coordinates that is visited
by the particle is divided into different zones and the time that the
particle stays in the zone before leaving is one instance of the waiting
time. The choice of the width of the zone is somewhat arbitrary since there
is no dynamics dependent action scale which is applicable to all of phase
space. For example, the resonance width is not relevant at small or large
amplitudes and if there were multiple resonances, there would be multiple
widths. We therefore calculated the waiting time distribution twice, once
with a chosen width such that there was enough statistics in each zone
and the second time with twice the width. In most cases we found that the
parameters of the distribution change by less than 10\%; we take this to be
a sign of convergence of the distribution. We find that the exponential
function is not a good fit to the distribution for either resonance. The
results for a fit to a power law distribution are shown in Figure 
\ref{fig: waiting_dist}. The distributions are plotted on a log-log scale
for several initial amplitudes where there is significant amplitude
growth. On the $2(3\nu_x-2\nu_s)=2$ resonance, most of the points (with the
exception of the single occurrence events with long waiting 
times) lie on
straight lines showing that a power law is a reasonable fit. The 
power law exponents for the different amplitudes are close. For the
amplitudes shown in this figure, the waiting law distributions are
\beq
 w(t) \sim t^{-\alp}, \;\;\;\; 2.4 \le \alp \le 2.7
\eeq
On the $4\nu_x - 2\nu_s = 1$ resonance, the waiting time distribution can
also be fit by a power law distribution but the range of variation in the
exponent $\alp$ is larger: $1.4 \le \alp \le 2.7$. The greater variability
in the exponent is expected to have an impact of the dependence of the
diffusion rate at different amplitudes on this resonance. 

\section{Fractional diffusion equation}

Since the waiting time distribution suggests that the transport near
resonances is non-Markovian, we need to establish an alternative to the regular 
diffusion equation. 
For a Markov process, the regular diffusion equation is obtained from
the Chapman-Kolmogorov master equation, a derivation is sketched in Appendix
A. In Appendix A we also derive a different master equation using general 
jump size and
waiting time distributions for a CTRW process in action space
following a method outlined in \cite{Sokolov}. The master equation for the
density in action angle space that we obtain is
\beq
\frac{\del }{\del t}\rho({\bf J},{\bf \theta}) \!\! =  \!\! \frac{1}{\tau}
\int \int d\Dl {\bf J} d\Dl {\bf \theta}
\Psi({\bf J} - \Dl{\bf J},{\bf \theta} - \Dl{\bf \theta};
\Dl {\bf J}, \Dl{\bf \theta})L_t \rho({\bf J} - \Dl{\bf J},{\bf \theta} - \Dl{\bf \theta},t) - \frac{1}{\tau} L_t \rho({\bf J},{\bf \theta},t) 
\eeq
where $L_t$ is an integral operator given by
\beq
\frac{1}{\tau} L_t \rho({\bf J},{\bf \theta};t) =
{\cal L}^{-1}\left[ 
\frac{s \hat{w}(s;{\bf J},{\bf \theta})}{1 - \hat{w}(s;{\bf J},{\bf \theta})} \hat{\rho}({\bf J},{\bf \theta},s) \right]
\eeq
Here ${\cal L}^{-1}$ is an inverse Laplace transform, $\tau$ is a 
time parameter in the waiting time distribution $w(t,{\bf J})$, 
$\hat{w}(s;{\bf J}), \hat{\rho}(s;{\bf J})$ are the Laplace
transforms of the waiting time distribution and the density respectively.
In the Appendix we then show that expanding this master equation in a 
Taylor series in the same manner as is done for the Chapman-Kolmogorov
equation, the following fractional diffusion equation is obtained for a
power law waiting time distribution $w(t,{\bf J}) \sim t^{-\alp({\bf J})}$
\beq
\frac{\del\rho}{\del t} 
 = \sum_k \sum_l \frac{\del}{\del J_k}[D_{kl} \frac{\del}{\del J_l}] 
\frac{1}{\Gm(1-\alp({\bf J}))}\left[ \frac{\del}{\del t}\int_0^t dt' 
\frac{\rho({\bf J},t')}{(t-t')^{\alp({\bf J})}}
\right]
\eeq
Here the exponent $\alp$ depends on the action ${\bf J}$ which will be true
in general and $D_{kl}$ are action dependent diffusion coefficients,
defined in the appendix. It remains to be verified that this fractional
diffusion equation describes the dynamics near resonances, as seen in the
particle tracking simulations. However, this diffusion equation has been derived
under general considerations of a CTRW process which the dynamics near the
SBR resonance appears to follow. 
Given the large variations in the diffusion
coefficients, the solution of this diffusion equation will likely require a 
special purpose numerical algorithm. The density can then be used to 
calculate the beam lifetime and various moments such as the emittance .

\section{Discussion}

We have studied the detailed transport process near two low order horizontal
synchro-betatron resonances driven by beam-beam interactions at a crossing
angle. We found that the horizontal beam profiles develop long beam tails. 
The horizontal beam distribution evolves from an initially Gaussian 
distribution to a
Levy stable distributions on both resonances. The Levy stable distributions
are solutions of simple fractional diffusion equations which describe some
anomalous diffusion processes. The evolution of the variance in action at
several initial values characterizes the nature of the diffusion in phase
space. At small amplitudes there is no diffusion, then there is a narrow
region where the motion is super-diffusive (the variance grows faster than
linearly with time), followed by a broad region where the motion is 
sub-diffusive (the variance grows slower than linearly with time) and 
finally no diffusion at large amplitudes. The width and the location of the 
super-diffusive region depends on the resonance, the width is narrower for
the weaker resonance. This super-diffusive region is also marked by
signatures of bounded chaos and particles do not experience large amplitude
growth. For both resonances, this region is located at the lower edge of
the resonance islands. The broad sub-diffusive region abuts the 
super-diffusive region and continues until about 5-6$\sg$ depending on the
resonance. Here particles do migrate to larger amplitudes. We do not observe
regular diffusion anywhere in phase space on either resonance with the particle
distributions we used. 

The jump size distribution and the waiting time distribution, key ingredients
of a continuous time random walk process, were found by analysis of single
particle tracking data. The jump size distributions for both resonances were 
similar - in the linear regions of phase space, the distributions in $\Dl x$ 
are close to the 
arcsine distribution while in the nonlinear regions they have a more 
complex shape. The similarity of these distributions for the two resonances
suggests that these may be universal features near such resonances. When the
waiting time distributions follows an exponential law, the stochastic process
is Markovian. We find that the waiting time distribution follows instead
a power law, again for both resonances. Since the process is non-Markovian,
the regular diffusion equation cannot be used to describe the evolution of
the density. For a general CTRW process, we derived a master equation in
action-angle space which is applicable to processes with arbitrary jump
size and waiting time distributions. A fractional diffusion equation was
derived from this master equation. Numerical solutions of this diffusion
equation will allow computations of beam observables such as lifetimes and
emittance growth. 

This model can be tested against beam observations when anomalous diffusion is
suspected. Comparison of beam profiles with Levy stable distributions would be
a first check. Another indicator would be if the emittance of pencil
beams grow nonlinearly with time. This could then be followed by
measurements of diffusion coefficients at different amplitudes, using them
in the fractional diffusion equation and comparing the numerically calculated
emittance growth and beam lifetime with the measured values.  

In this article we considered low order synchro-betatron resonances so as
to observe effects on a short time scale. Based on comparisons of the
two resonances studied here, we expect that the physics at high order
resonances (and hence more applicable to operational accelerators) will be
similar but on longer time scales. When multiple such resonances are present
simultaneously, the diffusion is likely to be anomalous but the phase
space dynamics will be more complicated. 
It is possible that the physics 
near space charge driven resonances may be similar to that obtained here but
that remains to be investigated. 

\vspace{2em}

\bec
{\bf \large Acknowledgments}
\eec

Fermilab is operated by Fermi Research Alliance, LLC under Contract 
No. DE-AC02-07CH11359 with the United States Department of Energy.

\clearpage

\appendix
\newcommand{\appsection}[1]{\let\oldthesection\thesection
\renewcommand{\thesection}{Appendix \oldthesection}
\section{#1}\let\thesection\oldthesection}

\section{Appendix: Regular and fractional diffusion equations}

\setcounter{equation}{0}
\renewcommand{\theequation}{\thesection.\arabic{equation}}

We briefly summarize the derivation of the diffusion equation in action-angle
space. We assume a Hamiltonian description $H({\bf J}, {\bf \theta})$ 
which has been perturbed from an integrable Hamiltonian $H_0({\bf J})$.
Let $\Psi({\bf J},{\bf \theta};\Dl {\bf J},\Dl {\bf \theta}$ be the 
transition probability for the action-angle variables to change from 
$({\bf J},{\bf \theta})$ to 
$({\bf J}+\Dl{\bf J},{\bf \theta}+\Dl{\bf \theta})$ in time $\Dl t$. 
The first major assumption is that the dynamics is Markovian.
For a Markov process, the particle density distribution at time $t+\Dl t$ only 
depends 
on its instantaneous state at $t$ and is independent of its previous history
provided $\Dl t$ is longer than a characteristic time $\tau$. 
Under this assumption, the density $\rho({\bf J},{\bf \theta},t)$
at time $t+\Dl t$ can be found by summing over all possible transitions
in time $\Dl t$. This results in the 
Chapman-Kolmogorov equation for the density  
\beq
\rho({\bf J},{\bf \theta},t+\Dl t) = \int \int
\rho({\bf J}-\Dl{\bf J},{\bf \theta}-\Dl{\bf \theta},t)
\Psi({\bf J}-\Dl{\bf J},{\bf \theta}-\Dl{\bf \theta};\Dl{\bf J},
\Dl{\bf \theta}) d(\Dl{\bf J}) d (\Dl{\bf \theta})
\label{eq: ChapKolm}
\eeq
Here $\Psi$ is the transition probability of jumps 
$(\Dl{\bf J},\Dl{\bf \theta})$.
Further assumptions need to be made including i)the angles evolve on
a faster time scale than the actions and their correlation decays 
rapidly, ii) the density in the long time limit is independent of the 
angle
iii) the transition probability can be factorized in the form 
$ \Psi({\bf J},{\bf \theta};\Dl{\bf J}, \Dl{\bf \theta}) = 
\Psi_J({\bf J}; \Dl {\bf J})\dl(\Dl{\bf \theta}- \dot{{\bf \theta}}\Dl t)$
iv) the changes in action and angle $\Dl{\bf J},\Dl {\bf \theta}$
are small during a time interval $\Dl t$.
Expanding the LHS and the RHS of
Equation (\ref{eq: ChapKolm}), keeping up to second order terms and then
taking the limit $\Dl t \rarw 0$, we obtain the Fokker-Planck equation
\beq
\frac{\del\rho}{\del t} 
 = -\nabla_{\bf J}\cdot[{\bf A}\rho]
 + \sum_{k}\sum_l \frac{\del^2 }{\del J_k \del J_l } [D_{kl} \rho]
\eeq
where the drift ${\bf A}$ and diffusion coefficients ${\bf D}$ are defined as
\beq
{\bf A}({\bf J}) = \lim_{\Dl{\bf J}\rarw 0, \Dl t\rarw 0}
 \frac{\lan \Dl {\bf J} \ran}{\Dl t}, \;\;\;
D_{kl}({\bf J}) = \lim_{\Dl{\bf J}\rarw 0, \Dl t\rarw 0} \half \frac{\lan \Dl J_k \Dl J_l \ran}{\Dl t}, \;\;\; \lan \Dl {\bf J} \ran \equiv \int \Dl {\bf J} 
\; \Psi_J({\bf J};\Dl {\bf J}) d {\bf J}
\eeq
Here $\Dl t$ is understood as a time shorter than a time scale over which
the density distribution evolves but longer than the time over which angle
correlations decay.

For Hamiltonian systems, there is a relation between the drift coefficient 
and the diffusion coefficients \cite{LichtLieb, VanKampen}
\beq
A_k = \half \sum_l\frac{\del}{\del J_l} D_{kl}
\label{eq: drift_diffcoef}
\eeq
then the Fokker-Planck equation simplifies to the diffusion equation
\beq
\frac{\del\rho}{\del t} 
 = \sum_k \sum_l \frac{\del}{\del J_k}[D_{kl} \frac{\del\rho}{\del J_l}]
\eeq
The assumptions of Markovian behavior and the smallness of the changes
in action-angle variables are crucial for the validity of this regular
diffusion equation. If these assumptions are invalid, then
this diffusion equation may not be the right model for the density evolution.

We now consider a more general master equation for a CTRW process in
action angle space with arbitrary jump size and waiting time distributions.
We use a method outlined in \cite{Sokolov}. 
It uses two basic balance conditions: the first states that a
change of density arises from the difference in the incoming flux 
$\Gm^+({\bf J},{\bf \theta},t)$ and the outgoing flux 
$\Gm^-({\bf J},{\bf \theta},t)$. 
\beq
\frac{\del }{\del t}\rho({\bf J},{\bf \theta}) = 
\Gm^+({\bf J},{\bf \theta},t) - \Gm^-({\bf J},{\bf \theta},t)
\label{eq: bal_1}
\eeq
The second balance condition states that the influx is composed of the 
outflux of particles from
all other phase space locations to that location
\beq
\Gm^+({\bf J},{\bf \theta},t) = \int \int d \Dl{\bf J} \; d\Dl{\bf \theta}
\; \Psi({\bf J} - \Dl{\bf J},{\bf \theta} - \Dl{\bf \theta};
\Dl {\bf J}, \Dl{\bf \theta})\Gm^-({\bf J}-\Dl {\bf J},{\bf \theta}-
\Dl{\bf \theta},t)  
\label{eq: bal_2}
\eeq
The outflux at $({\bf J},{\bf \theta},t)$ has contributions from particles 
that were present initially but left after waiting for time $t$ and those 
that arrived later before leaving 
\beq
\Gm^-({\bf J},{\bf \theta},t) = w(t;{\bf J},{\bf \theta})
\rho({\bf J},{\bf \theta},0) +
\int_0^t w(t-t';{\bf J},{\bf \theta}) \Gm^+({\bf J},{\bf \theta},t') dt'
\label{eq: bal_3}
\eeq
Substituting Eq. (\ref{eq: bal_3}) in Eq. (\ref{eq: bal_1}) and taking the 
Laplace transform, we obtain for the outflux
\beq
\Gm^-({\bf J},{\bf \theta},t) = {\cal L}^{-1}\left[ 
\frac{s \hat{w}(s;{\bf J},{\bf \theta})}{1 - \hat{w}(s;{\bf J},{\bf \theta})} \hat{\rho}({\bf J},{\bf \theta},s) \right]
\equiv \frac{1}{\tau} L_t \rho({\bf J},{\bf \theta};t)
\eeq
Here $\hat{w}(s;{\bf J},{\bf \theta})$ and 
$\hat{\rho}({\bf J},{\bf \theta},s)$ are the Laplace transforms in $s$ space,
$\tau$ is a relevant time parameter in the waiting time distribution.
and ${\cal L}^{-1}$ is the inverse Laplace transform. 
The last equality in this equation defines the integral operator $L_t$.
Substituting this back in Eq. (\ref{eq: bal_1}) and using 
Eq.(\ref{eq: bal_2}) we obtain
\beq
\frac{\del }{\del t}\rho({\bf J},{\bf \theta})  \!\! =  \!\! \frac{1}{\tau}
\int \int d\Dl {\bf J} d\Dl {\bf \theta}
\Psi({\bf J} - \Dl{\bf J},{\bf \theta} - \Dl{\bf \theta};
\Dl {\bf J}, \Dl{\bf \theta})L_t \rho({\bf J} - \Dl{\bf J},{\bf \theta} - \Dl{\bf \theta},t) - \frac{1}{\tau} L_t \rho({\bf J},{\bf \theta},t) 
\label{eq: mod_master}
\eeq
This is the modified master equation for the density. 

Now we derive the modified diffusion equation from this master equation.
We expand the RHS of Eq.(\ref{eq: mod_master})
in a Taylor series and keep up to 
second order terms. As before we define the coefficients 
\beq
{\bf A}({\bf J}) = \lim_{\Dl{\bf J}\rarw 0, }
 \frac{\lan \Dl {\bf J} \ran}{\tau}, \;\;\;
D_{kl}({\bf J}) = \lim_{\Dl{\bf J}\rarw 0} \half \frac{\lan \Dl J_k \Dl J_l \ran}{\tau}
\eeq
We assume that the same relation as in Eq. (\ref{eq: drift_diffcoef}) between the drift and diffusion coefficients holds. Then we have as the modified 
diffusion equation
\beq
\frac{\del\rho}{\del t} 
 = \frac{1}{\tau} \sum_k \sum_l \frac{\del}{\del J_k}[D_{kl} \frac{\del}{\del J_l}] L_t\rho
\label{eq: mod_diff}
\eeq
In cases where $(1/\tau)L_t\rho = \rho$, this is the regular diffusion 
equation.

Consider now two examples of a waiting time distribution, first 
an exponential waiting time
\beq 
w(t) = \frac{1}{\tau} \exp[-\frac{t}{\tau}],\;\;\;
 \Rarw \hat{w}(s) = \frac{1}{\tau} (\frac{1}{s + 1/\tau})
\eeq
The integral operator simplifies to 
\beq 
\frac{1}{\tau}L_t\rho = {\cal L}^{-1}[\frac{s \hat{w}(s)}{1 - \hat{w}(s)} 
\hat{\rho}({\bf J},s) ] = \rho({\bf J},t) 
\eeq
i.e. the modified diffusion equation reduces to the regular diffusion equation.

Now consider a power law waiting time
\beq 
w(t;{\bf J}) = \frac{1}{\tau}(\frac{t}{\tau})^{-\alp({\bf J})} 
\eeq
Here we let the exponent $\alp$ be action dependent. 
In the long time limit $t \rarw \infty$ or equivalently $ s \rarw 0$,  
\beq 
\frac{1}{\tau}L_t\rho = \Gm(1 -\alp({\bf J})) {\cal L}^{-1} [s^{\alp({\bf J})}
\hat{\rho}({\bf J},s)]
  = \mbox{}_0D_t^{\alp({\bf J})} \rho({\bf J},t) 
 \eeq
Here $\Gm$ is the Gamma function and 
$\mbox{}_0 D_t^{\alp({\bf J})}$ is a Riemann-Liouville fractional
derivative in time defined below. The diffusion equation for $\rho$ is
\beqr
\frac{\del\rho}{\del t} 
 & = & \sum_k \sum_l \frac{\del}{\del J_k}[D_{kl} \frac{\del}{\del J_l}] 
 \; \mbox{}_0 D_t^{\alp({\bf J})} \rho({\bf J},t)
 \nonumber \\
 & = & \sum_k \sum_l \frac{\del}{\del J_k}[D_{kl} \frac{\del}{\del J_l}] 
\frac{1}{\Gm(1-\alp({\bf J}))}\left[ \frac{\del}{\del t}\int_0^t dt' 
\frac{\rho({\bf J},t')}{(t-t')^{\alp({\bf J})}} \right] 
\label{eq: frac_diff}
\eeqr
This is a non-local in time (due to the waiting time distribution) 
integro-differential diffusion equation for the density.

\end{document}